\documentclass[12pt]{article}
\usepackage{graphicx}
\usepackage{amsmath}
\usepackage{bm}
\usepackage{colordvi}
\usepackage{color}
\usepackage{epsfig}
\usepackage{fancybox}

\newcommand\pubnumber{ANL-HEP-CP-12-64}
\newcommand\pubdate{March 11, 2013}

\def\argonne{HEP Division, Argonne National Laboratory\\ 9700 South Cass
Avenue, Argonne, Illinois 60439 USA} 

\def\Title#1{\begin{center} {\Large #1 } \end{center}}
\def\Author#1{\begin{center}{ \sc #1} \end{center}}
\def\Address#1{\begin{center}{ \it #1} \end{center}}

\newcommand\pubblock{\rightline{\begin{tabular}{l} \pubnumber\\
         \pubdate  \end{tabular}}}
\newenvironment{Abstract}{\begin{quotation}  }{\end{quotation}}
\newenvironment{Presented}{\begin{quotation} \begin{center} 
             PRESENTED AT\end{center}\bigskip 
      \begin{center}\begin{large}}{\end{large}\end{center} \end{quotation}}
\def\Acknowledgements{\bigskip  \bigskip \begin{center} \begin{large}
             \bf ACKNOWLEDGEMENTS \end{large}\end{center}}

\def\beq{\begin{equation}}
\def\eeq#1{\label{#1}\end{equation}}
\def\eeqn{\end{equation}}

\def\beqa{\begin{eqnarray}}
\def\eeqa#1{\label{#1}\end{eqnarray}}
\def\eeqan{\end{eqnarray}}

\let\bar=\overbar

\def\Dslash{\not{\hbox{\kern-4pt $D$}}}
\def\dslash{\not{\hbox{\kern-2pt $\del$}}}

\def\msb{{\bar{\ssstyle M \kern -1pt S}}}

\begin{document}
\begin{titlepage}
\pubblock

\vfill
\Title{Theory of Charmonium Production}
\vfill
\Author{Geoffrey T.\ Bodwin}
\Address{\argonne}
\vfill
\begin{Abstract}
I give an overview of the current status of the theory of
charmonium production in hard-scattering processes and summarize
the present state of comparisons between theory and experiment. 
\end{Abstract}
\vfill
\begin{Presented}
Charm 2012\\ The 5th International Workshop on Charm Physics\\
Honolulu,Hawaii, May 14--17, 2012
\end{Presented}
\vfill
\end{titlepage}
\def\thefootnote{\fnsymbol{footnote}}
\setcounter{footnote}{0}

\section{Introduction}

Since their discovery in the mid-1970s, heavy-quarkonium states have
played an important role in helping us to understand QCD. Because
heavy-quarkonium states are nonrelativistic, they allow the application
of theoretical tools that simplify and constrain the analyses of
nonperturbative effects. Hence, heavy-quarkonium states provide a
unique laboratory in which to explore the interplay between perturbative
and nonperturbative effects in QCD. 

In this talk I begin by describing the theoretical framework for
calculations of inclusive quarkonium production processes, with emphasis
on the NRQCD factorization approach \cite{Bodwin:1994jh}. I then
describe the current status of the comparison between theoretical
predictions and experimental measurements.\footnote{See
Refs.~\cite{Brambilla:2004wf,Brambilla:2010cs} for some further details
regarding these issues.}

\section{Theoretical Framework}

A number of theoretical approaches have been proposed for the
calculation of heavy-quarkonium production processes. These include the
NRQCD factorization approach \cite{Bodwin:1994jh}, the fragmentation
approach \cite{Kang:2011zza,Kang:2011mg}, the color-singlet model (CSM)
\cite{Kartvelishvili:1978id,Chang:1979nn,Berger:1980ni,Baier:1981uk,Baier:1983va},
the color-evaporation model (CEM)
\cite{Fritzsch:1977ay,Halzen:1977rs,Amundson:1995em,Amundson:1996qr},
and the $k_T$-factorization approach
\cite{Yuan:2000qe,Baranov:2002cf,Baranov:2007ay,Baranov:2007dw}. Only
NRQCD factorization and the fragmentation approach are believed to be
methods that could be derived from QCD. NRQCD factorization is the
default model for most current studies of quarkonium production.

\subsection{NRQCD Factorization of the Inclusive Production Cross Section}

Nonrelativistic QCD (NRQCD) is an effective field theory that describes
the behavior of bound states of a heavy-quark ($Q$) and a
heavy-antiquark($\bar Q$) when the velocity $v$ of the $Q$ or $\bar Q$
in the $Q\bar Q$ rest frame is nonrelativistic ($v\ll 1$). Some years
ago, Bodwin, Braaten and Lepage \cite{Bodwin:1994jh} conjectured that
the inclusive cross section for producing a quarkonium at large momentum
transfer ($p_T$ or $p^*$) can be written as a sum of products of 
``short-distance'' coefficients times NRQCD matrix elements:
\begin{equation}
\sigma(H)=\sum_n F_n(\Lambda)\langle 0|
{\cal O}_n^H(\Lambda)|0\rangle.
\label{NRQCD-fact}
\end{equation}
The ``short-distance'' coefficients $F_n(\Lambda)$ are essentially the
process-dependent partonic cross sections to make a $Q\bar Q$ pair,
convolved with the parton distributions of the incoming hadrons. The
NRQCD long-distance matrix elements (LDMEs) $\langle 0|{\cal
O}_n^H(\Lambda)|0\rangle$ are the probability for a $Q\bar Q$ pair to
evolve into a heavy quarkonium.  $\Lambda$ is the factorization scale,
which is the cutoff of the effective field theory. The NRQCD matrix
elements are vacuum expectation values of four-fermion operators in
NRQCD, but with a projection onto an intermediate state of the
quarkonium $H$ plus anything:
\begin{equation}
\langle 0|{\cal O}_n^H(\Lambda)|0\rangle=\langle 0|\chi^\dagger \kappa_n\psi 
\biggl(\sum_X |H+X\rangle\langle H+X|\biggr) \psi^\dagger 
\kappa'_n\chi|0\rangle.
\end{equation}
Here, $\psi^\dagger$ and $\chi$ are two-component (Pauli) fields that
create a heavy quark and a heavy antiquark, respectively, and $\kappa_n$
and $\kappa_n'$ are direct products of Pauli and color
matrices.\footnote{It was pointed out by Nayak, Qiu, and Sterman that
gauge invariance requires that the definitions of the NRQCD LDMEs 
include Wilson lines that run from the quark and antiquark
fields to infinity \cite{Nayak:2005rw}. For simplicity, I have omitted
these Wilson lines here.}

The short-distance coefficients have expansions in powers of
$\alpha_s$. The LDMEs are nonperturbative, but they
are conjectured to be universal, {\it i.e.,} process independent.
Only the color-singlet production LDMEs are simply related to the decay 
LDMEs. The LDMEs have a known scaling with $v$, where 
$v^2\approx 0.23$ for the $J/\psi$ and $v^2\approx 0.1$ for the 
$\Upsilon(1S)$. Hence, the NRQCD factorization formula is a double 
expansion in powers of $\alpha_s$ and $v$.

The current phenomenology of $J/\psi$, $\psi(2S)$, and
$\Upsilon(nS)$ production uses LDMEs through relative order
$v^4$:
\begin{equation}
\begin{tabular}{ll}
$\langle{\cal O}^{H}(^3S_1^{[1]})\rangle$ &($O(v^0)$),\\
$\langle{\cal O}^{H}(^1S_0^{[8]})\rangle$ &($O(v^3)$),\\
$\langle{\cal O}^{H}(^3S_1^{[8]})\rangle$ &($O(v^4)$),\\
$\langle{\cal O}^{H}(^3P_J^{[8]})\rangle$ &($O(v^4)$).
\end{tabular}
\label{S-wave-MEs}
\end{equation}
Here, the superscripts $H$ stand for any spin-triplet, $S$-wave
quarkonium state, including the $J/\psi$, the $\psi(2S)$, and the
$\Upsilon(nS)$ states, all of which have different NRQCD matrix elements.
The quantities in parentheses are the spin, orbital-angular-momentum,
and color (singlet or octet) quantum numbers of the $Q\bar Q$ pair that
evolves into the quarkonium state $H$. 

A key feature of NRQCD factorization is that quarkonium production can
occur through color-octet, as well as color-singlet, $Q\bar Q$ states.
The color-singlet production LDMEs $\langle{\cal
O}^{H}(^3S_1^{[1]})\rangle$ can be determined from quarkonium
electromagnetic decay rates (up to corrections of order $v^4$). However,
the color-octet LDMEs must be determined through comparisons of
theoretical predictions with measurements. Once they have been
fixed through comparisons between theory and experiment in one or
more processes, the LDMEs can be used, by virtue of the
property of universality, to make predictions for other processes.

One obtains the CSM by retaining, for a given process, only the
contribution that is associated with the color-singlet LDME of the
lowest nontrivial order in $v$. The CSM is theoretically inconsistent in
that it leads to uncanceled infrared divergences in calculations of
production and decay processes for $P$-wave quarkonium states.

\subsubsection{Status of a Proof of NRQCD Factorization}

A proof of NRQCD factorization is complicated because gluons can dress
the quarkonium basic production process in ways that apparently violate
factorization in individual Feynman diagrams. A proof of factorization
would involve a demonstration that diagrams in each order in $\alpha_s$
can be re-organized so that (1) all soft singularities cancel or can be
absorbed into NRQCD LDMEs and (2) all collinear singularities and
spectator interactions can be absorbed into parton distributions. Nayak,
Qiu, and Sterman have demonstrated factorization for the leading power in
$1/p_T$ through next-to-next-to-leading order (NNLO) in $\alpha_s$
(Refs.~\cite{Nayak:2005rw,Nayak:2005rt,Nayak:2006fm}). However, it is not
known if the proof generalizes to all orders in $\alpha_s$. Note that an
all-orders proof is essential because potential violations of
factorization involve soft gluons, for which $\alpha_s$ is not a good
expansion parameter. It seems likely that, if a proof of NRQCD
factorization can be established, it would hold only for values of $p_T$
that are greater than the heavy-quark mass $m_Q$.

\subsection{The Problem of Large Higher-Order Corrections}

\subsubsection{Higher-Order Corrections in $\bm{J/\psi}$ and 
$\bm{\Upsilon(1S)}$ Production}

\begin{figure}[h!tb]
\begin{center}
\includegraphics[width=0.45\columnwidth]{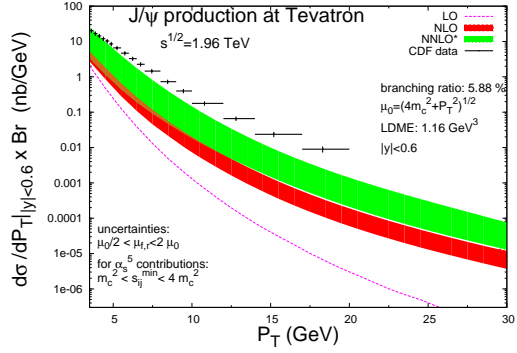}
\end{center}
\caption{Calculations of the $J/\psi$ production cross section in the
color-singlet channel compared to CDF Run~II data
\cite{Abulencia:2007us}. LO: leading order in $\alpha_s$. NLO:
next-to-leading-order in $\alpha_s$. NNLO$^*$: next-to-next-to-leading
order in $\alpha_s$, real corrections only. (Figure from P.~Artoisenet,
based on work by P.~Artoisenet, J.~Campbell, J.P.~Lansberg, F.~Maltoni,
and F.~Tramontano \cite{Artoisenet:2008fc}.)}
\label{psi-color-singlet-HO}
\end{figure}

When one calculates quarkonium production cross sections differential in
$p_T$ in the NRQCD factorization approach, large corrections appear in
higher orders in perturbation theory. These large corrections were first
noticed in calculations of quarkonium production at hadron-hadron
colliders through the color-singlet channel \cite{Artoisenet:2008fc}. An
example of this phenomenon is shown in Fig.~\ref{psi-color-singlet-HO}.
The curve labeled ``NNLO$^*$'' is an estimate that is based on only the
real-emission contributions at NNLO in $\alpha_s$. Note that the
color-singlet contributions alone do not explain the data.

\begin{figure}[h!tb]                 
\begin{center}
\includegraphics[width=0.45\columnwidth]{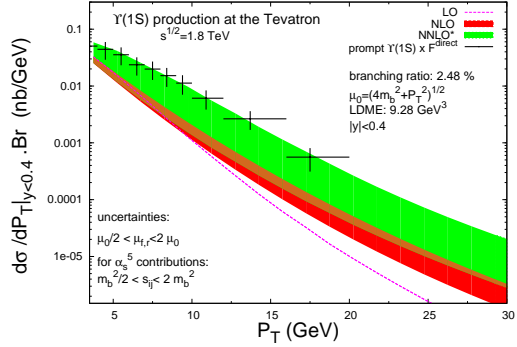}
\end{center}
\caption{Calculations of the $\Upsilon(1S)$ production cross section in the
color-singlet channel compared to CDF Run~II data \cite{Acosta:2001gv}.
LO: leading order in $\alpha_s$. NLO: next-to-leading-order in
$\alpha_s$. NNLO$^*$: next-to-next-to-leading order in $\alpha_s$, real
corrections only. (Figure from P.~Artoisenet, based on work by
P.~Artoisenet, J.~Campbell, J.P.~Lansberg, F.~Maltoni, and F.~Tramontano
Ref.~\cite{Artoisenet:2008fc}.)}
\label{ups-color-singlet-HO}                                             
\end{figure}

An example of large higher-order corrections in $\Upsilon(1S)$ production in
the color-singlet channel is shown in Fig.~\ref{ups-color-singlet-HO}.
The next-to-leading order (NLO) results that are shown in this figure
were confirmed by Gong and Wang \cite{Gong:2008sn}. In this case, the
data could be explained by NNLO$^*$ color-singlet production alone.
However, given the large error bars, there is still room for a
substantial of color-octet contribution.

A very large correction in next-to-leading order in $\alpha_s$ also
arises in $J/\psi$ production through the color-octet ${}^3P_J$ channel,
where a negative correction has been found
\cite{Ma:2010yw,Butenschoen:2010rq}. However, NLO corrections to the
$S$-wave color-octet channels are small \cite{Gong:2008ft}.  The
corrections for production at the Tevatron are approximately factors
of $1.235$ for the ${}^1S_0$ color-octet channel and $1.139$ for the
${}^3S_1$ color-octet channel.

These results raise several questions: (1) Does the perturbation series
converge? (2) How does one understand the different sizes of the
higher-order corrections for different channels? 

\subsubsection{Explanation of the Large Corrections}

It has been suggested by Campbell, Maltoni, and Tramontano 
\cite{Campbell:2007ws} that the large
higher-order corrections to quarkonium production can be understood as
follows: At high $p_T$, higher powers of $\alpha_s$ can be offset by a
less rapid fall-off with $p_T$. For example, for production of the
$J/\psi$ through the $S$-wave, color-singlet channel, the contribution
at leading order (LO) in $\alpha_s$ goes as $\alpha_s^3 m_c^4/p_T^8$. In
contrast, the NLO contribution goes as $\alpha_s^4 m_c^2/p_T^6$ and is
enhanced at high $p_T$ relative to the LO contribution.\footnote{The NLO
correction also contains a contribution, which arises from  charm-quark
fragmentation into a $J/\psi$, that goes as $\alpha_s^4/p_T^4$. This
charm-quark fragmentation contribution happens to be small numerically.}
The NNLO contribution to $S$-wave, color-singlet production of the $J/\psi$
is further enhanced at high $p_T$, since it goes as $\alpha_s^5/p_T^4$,
owing to contributions in which a gluon fragments into the $J/\psi$. The
leading power behavior of the cross section at high $p_T$ is $1/p_T^4$
\cite{Kang:2011zza,Kang:2011mg}. Hence, one expects no further large
kinematic enhancements beyond NNLO.

The NNLO calculation of  color-singlet, $S$-wave production of $S$-wave
charmonium states is, so far, incomplete, as only the real-emission
(NNLO$^*$) correction has been computed.  It has been noticed by Ma,
Wang, Chao \cite{Ma:2010jj} that the color-singlet NNLO$^*$ correction
seems be dominated by contributions that are proportional to
$\log^2(p_T^2/p_{T{\rm cut}}^2)$, where $p_{T{\rm cut}}^2>m_Q^2$ is an
artificial cutoff that is introduced to make the calculation finite.
These contributions will cancel when virtual corrections are included in
the complete NNLO calculation. Hence, the complete NNLO contribution is
likely to be significantly smaller than the NNLO$^*$ contribution and
cannot, by itself, explain the observed production cross sections of
$S$-wave charmonium states.

Kinematic enhancements also account for the large corrections in the
color-octet channels. The color-octet ${}^3S_1$ channel receives a small
correction in NLO because it already has $1/p_T^4$ behavior in LO, owing
to a gluon-fragmentation contribution. The color-octet ${}^3P_J$ channel
receives a large correction in NLO because it first shows $1/p_T^4$
behavior in NLO, again owing to gluon fragmentation. The color-octet
${}^1S_0$ channel also shows first shows $1/p_T^4$ behavior in NLO, and
that behavior arises from gluon fragmentation. However, the NLO
correction to the color-octet ${}^1S_0$ channel is numerically small at
moderate $p_T$ because the fragmentation process has little support near
$z=1$.\footnote{The absence of peaking near $z=1$ may reflect the fact
that the color-octet ${}^1S_0$ fragmentation process contains no soft
divergences in full QCD \cite{bodwin-lee-frag}. Such soft divergences
are ultimately absorbed into the NRQCD LDMEs, rendering the physical
cross section finite, but the remnants of the soft divergences can
produce a peaking near $z=1$.}

Note that, at NLO, all of the color-octet channels that appear in
$S$-wave quarkonium production already contain contributions that go as
$1/p_T^4$. This suggests that no further large enhancements will
occur in still higher orders in $\alpha_s$ in these channels. At
leading-order (LO) in $\alpha_s$, only the ${}^3S_1$ color-octet channel
has a $1/p_T^4$ behavior. This fact has been used to argue that the 
${}^3S_1$ color-octet  channel is dominant at large $p_T$. However, at
NLO, all of the color-octet channels have a $1/p_T^4$ behavior, and
so the argument for the dominance of the ${}^3S_1$ color-octet channel
at large $p_T$ is not valid.

\subsection{The Fragmentation Approach}

Kang, Qiu, and Sterman \cite{Kang:2011zza,Kang:2011mg} 
have suggested that one reorganize the
perturbation expansions for inclusive quarkonium production cross
sections according to the $p_T$ dependence of the various
contributions. Specifically, they find that the leading behavior in
$1/p_T$ in the production cross section ($1/p_T^4$) comes from
contributions in which single-parton production cross sections are
convolved with the fragmentation functions for a single parton into a
quarkonium:
\begin{equation}
d\hat\sigma_{A+B\to i+X}\otimes D_{i\to H}.
\end{equation}
They also find that the first subleading behavior         
in $1/p_T$ in the production cross section ($m_Q^2/p_T^6$) comes from 
$Q\bar Q$ production cross sections convolved with
fragmentation functions for a $Q\bar Q$ pair into a quarkonium:
\begin{equation}
d\hat\sigma_{A+B\to Q\bar Q+X}\otimes D_{Q\bar Q\to H}.
\end{equation}
Kang, Qiu, and Sterman have provided convincing arguments that these
results hold to all orders in perturbation theory, up to corrections of
order $m_Q^4/p_T^8$.

The fragmentation approach of Kang, Qiu, and Sterman is well suited to
the analysis of phenomenon of large higher-order corrections  because it
allows one to identify the specific sources of the kinematic
enhancements at high $p_T$, which arise from the leading and first
subleading powers in $1/p_T$ in the quarkonium production rates.
Hence, the fragmentation approach should allow one to reduce the
uncertainties in the theoretical predictions by focusing on those
processes that produce the large corrections, calculating them to yet
higher orders in $\alpha_s$, and resumming the associated
logarithms of $p_T^2/m_Q^2$, which are large at high $p_T$.

It is important to check that the fragmentation formalism really does
account for all of the large corrections. It has been confirmed that the
fragmentation contribution reproduces most of the large correction in
the color-singlet channel at NLO \cite{Kang:2011mg}. There is also an
estimate of the fragmentation contribution in the $^3P_J$ color-octet
channel that indicates that it gives a significant part of the NLO
correction in that channel \cite{bodwin-lee-frag}. Several other tests
of the fragmentation approach are in progress.

The values of the nonperturbative fragmentation functions that appear in
this formalism are, so far, unknown.  However, if NRQCD factorization
holds, then the fragmentation functions of Kang, Qiu, and Sterman can be
written as a sum of NRQCD LDMEs times perturbatively calculable
short-distance coefficients. In that case, the fragmentation approach is
equivalent to a reorganization of the NRQCD factorization formula
(\ref{NRQCD-fact}) according to the behaviors $1/p_T^4$ and
$m_Q^2/p_T^6$, with the stipulation that there are corrections of order
$m_Q^4/p_T^8$ that are not accounted for in the fragmentation approach.

\section{Comparisons of NRQCD Factorization with Experiment}

\subsection{Overview}

NLO corrections to charmonium production cross sections and
polarizations have been computed for many production processes: $J/\psi$
and $\psi(2S)$ production cross sections and polarization at the
Tevatron and the LHC;  $J/\psi$ and $\psi(2S)$ production cross sections
at RHIC; $J/\psi$ photoproduction cross sections and polarization at
HERA; the $J/\psi+\eta_c$ production cross section, the $J/\psi+c\bar c$
production cross section, and the $J/\psi+X(\hbox{non-$c\bar c$})$
production cross section in $e^+e^-$ annihilation at the $B$ factories.
An NLO calculation of the $\chi_J$ production cross sections for the
individual $J$ states also exists \cite{Ma:2010vd}.

Generally, data and the NLO predictions of NRQCD factorization for
quarkonium production agree, within errors.\footnote{See the global fit
of NRQCD matrix elements of Butensch\"on and Kniehl
\cite{Butenschoen:2011yh,Butenschoen:2012qh} and
Ref.~\cite{Brambilla:2010cs} for some further details.} There are three
significant exceptions: (1) quarkonium polarization in hadron-hadron
collisions , (2) the cross section for the production of
$J/\psi+X(\hbox{non-${\bm c\bar c}$})$ in $e^+e^-$ annihilation at
Belle, and (3) the cross section for $J/\psi$ production in
$\gamma\gamma$ scattering at LEP~II. I will discuss each of these
exceptions below, following a discussion of the extraction of the NRQCD
LDMEs from hadroproduction and photoproduction data.

\subsection{Extraction of NRQCD LDMEs at NLO}

I now describe an important issue that arises in extracting NRQCD matrix
elements from fits of NLO NRQCD predictions to $J/\psi$ production data.

Recently, two groups, Ma, Wang, and Chao \cite{Ma:2010yw,Ma:2010jj} and
Butensch\"on and Kniehl \cite{Butenschoen:2010rq}, have carried out the
first complete calculations of $J/\psi$ hadroproduction at NLO. These
calculations are complete in the sense that they include all of the
color-singlet and color-octet channels that contribute through order
$v^4$. [See Eq.~(\ref{S-wave-MEs}).] The results of the two groups for
the NRQCD short-distance coefficients agree. However, the fitted NRQCD
matrix elements are very different.

Using the CDF Run~II data, Ma, Wang, and Chao \cite{Ma:2010yw,Ma:2010jj}
could fit only two linear combinations of matrix elements unambiguously:
\begin{eqnarray}
M_{0,r_0}&=&\langle O^{\psi} \big(
^1S_0^{[8]}\big)\rangle+(r_0/m_c^2) \langle O^{\psi} \big(
^3P_0^{[8]}\big)\rangle=(7.4\pm 1.9)\times 
10^{-2}~\hbox{GeV}^3,\nonumber\\
M_{1,r_1}&=&\langle O^{\psi} \big(
^3S_1^{[8]}\big)\rangle+(r_1/m_c^2) \langle O^{\psi} \big(
^3P_0^{[8]}\big)\rangle=(0.05\pm 0.03)\times 10^{-2}~\hbox{GeV}^3,
\label{ma-wang-chao-fit}
\end{eqnarray}
where $r_0=3.9$ and $r_1=-0.56$ were chosen on the basis of approximate 
relations between the short-distance coefficients. 

Butensch\"on and Kniehl \cite{Butenschoen:2011yh} 
used their NLO calculations for $ep$, 
$\gamma\gamma$, and $e^+e^-$ production to fit all three color-octet 
LDMEs, using data from the Tevatron, LHC, RHIC, HERA, LEPII and KEKB:
\begin{eqnarray}
\langle O^{\psi} \big(^1S_0^{[8]}\big)\rangle&=&(4.76\pm 0.06)\times 
10^{-2}~\hbox{GeV}^3,\nonumber\\
\langle O^{\psi} \big(^3S_1^{[8]}\big)\rangle&=&(0.265\pm 0.014)\times 
10^{-2}~\hbox{GeV}^3,\nonumber\\
\langle O^{\psi} \big(^3P_0^{[8]}\big)\rangle/m_c^2&=&(-0.716\pm 0.089)\times 
10^{-2}~\hbox{GeV}^3,
\end{eqnarray}
which implies that
\begin{eqnarray}
M_{0,r_0}&=&(2.17\pm 0.56)\times 10^{-2}~\hbox{GeV}^3,\nonumber\\
M_{1,r_1}&=&(0.62\pm 0.08)\times 10^{-2}~\hbox{GeV}^3.
\end{eqnarray}

There are many small differences in the fitting procedures: (1) Ma,
Wang, and Chao included feeddown from the $\psi(2S)$ and the $\chi_{cJ}$
states in the theoretical prediction, but Butensch\"on and Kniehl did
not; (2) Ma, Wang, and Chao used a 2-parameter constrained fit, while
Butensch\"on and Kniehl used a 3-parameter fit; (3) Ma, Wang, and Chao
applied a cut $p_T> 7$~GeV to the CDF data, while Butensch\"on and
Kniehl applied a cut $p_T>3$~GeV to the CDF data. However, the most
important difference between these fits is the use of the low-$p_T$ H1
data by Butensch\"on and Kniehl. Indeed, the fit of Butensch\"on and
Kniehl to the Tevatron and HERA data alone \cite{Butenschoen:2010rq}
gives a very similar result to that of the global fit:
\begin{eqnarray}  
M_{0,r_0}&=&(2.5\pm 0.08)\times 10^{-2}~\hbox{GeV}^3,\nonumber\\
M_{1,r_1}&=&(0.59\pm 0.02)\times 10^{-2}~\hbox{GeV}^3.           
\end{eqnarray} 
Note that, for the HERA data, $p_T$ lies in the
range $1~\hbox{GeV}< p_T < 3~\hbox{GeV}$. It is not clear that NRQCD 
factorization, if proven, would hold at such low values of $p_T$.

\begin{figure}[h!tb]
\begin{center}
\includegraphics[width=0.45\columnwidth]{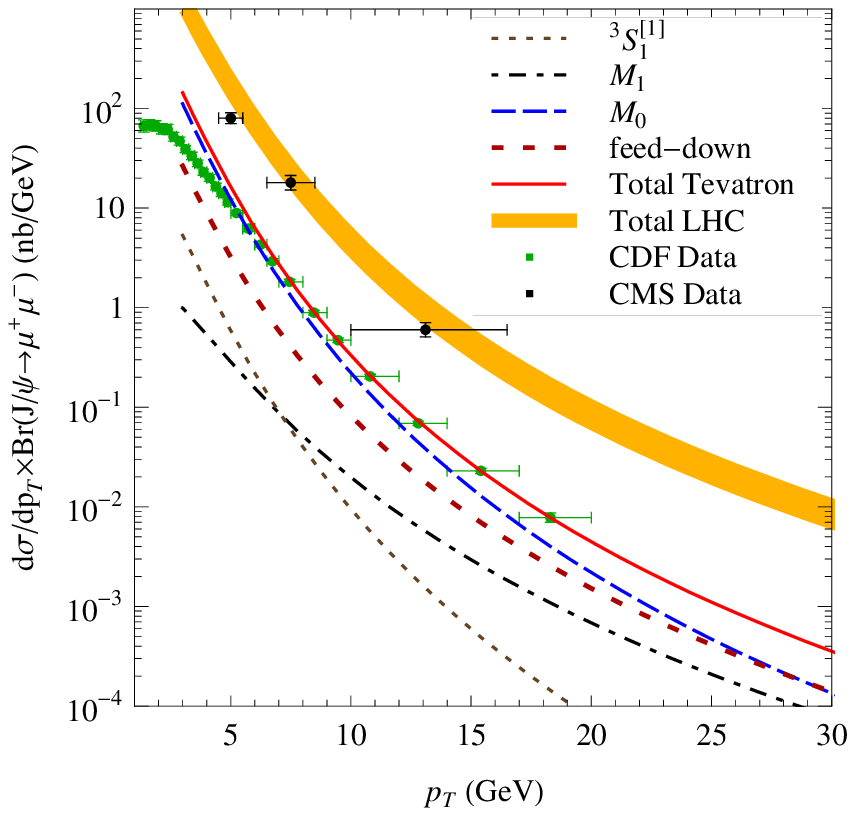}\qquad
\includegraphics[width=0.45\columnwidth]{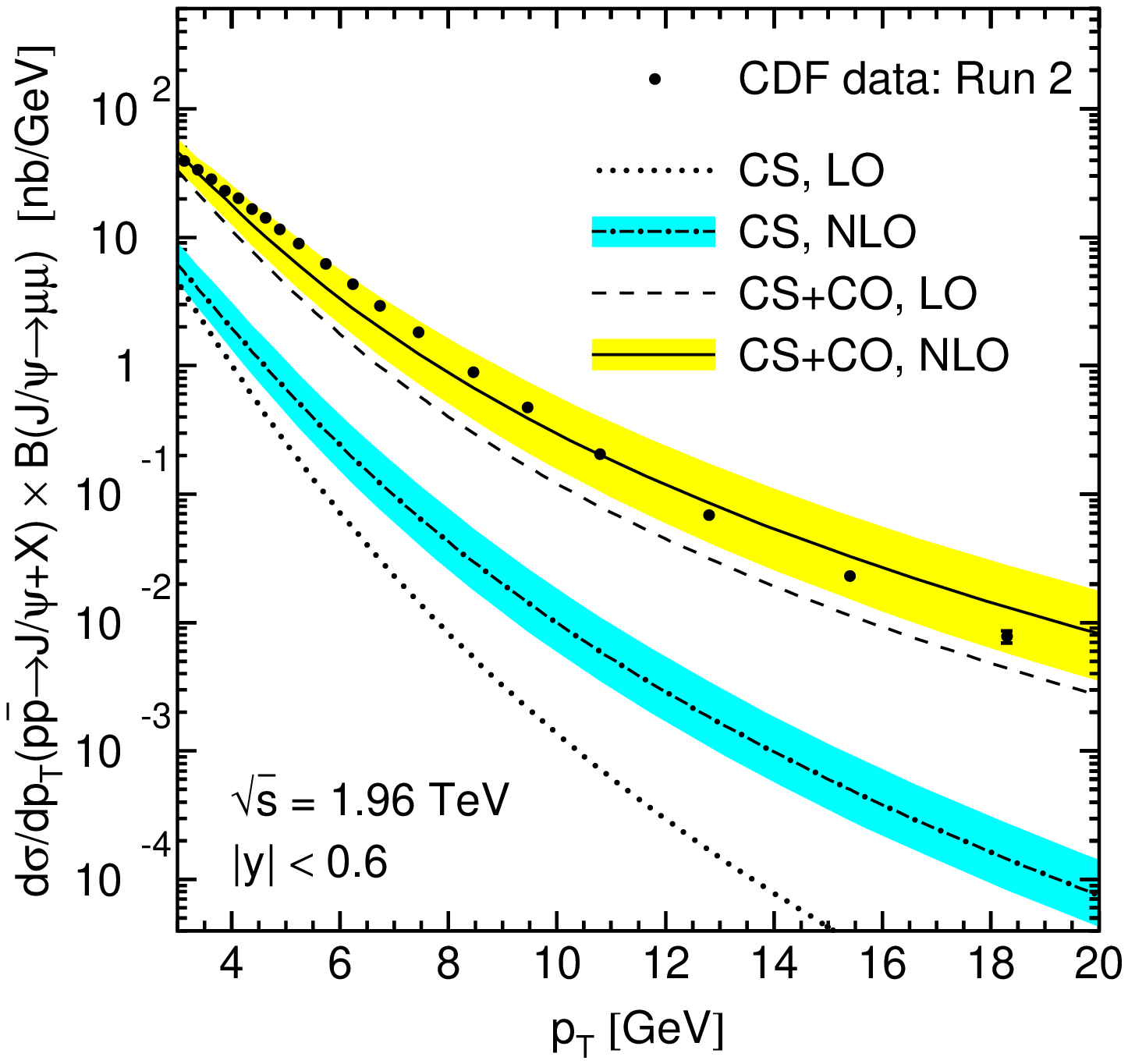}
\end{center}
\caption{Comparisons of NLO fits to the CDF data \cite{Acosta:2004yw}.
Left: The fit of Ma, Wang, and Chao \cite{Ma:2010yw,Ma:2010jj}. (Figure from
Ref.~\cite{Ma:2010yw}.) Right: The global fit of Butensch\"on and Kniehl
\cite{Butenschoen:2011yh}. In this, and subsequent, figures ``CS'' is 
the color-singlet contribution and ``CO'' is the color-octet 
contribution. (Figure from Ref.~\cite{Butenschoen:2011yh}.)
}
\label{ma-buten-cdf}
\end{figure}

Both fits describe the data within errors, but, as can be seen from
Fig.~\ref{ma-buten-cdf}, 
the shape of the Ma, Wang, and Chao fit agrees with
the CDF data better than the shape of the Butensch\"on and Kniehl fit.
The fit of Butensch\"on and Kniehl to the CDF data clearly overshoots
the data at large $p_T$, while the fit of Ma, Wang, and Chao does not.
Resummation of large logarithms of $p_T/m_c^2$ at large $p_T$ would
lower the theoretical predictions because the fragmentation process
softens the $J/\psi$ momentum.  This resummation correction would improve
the agreement of the fit of Butensch\"on and Kniehl with the CDF data and
worsen the agreement of the fit of  Ma, Wang, and Chao with the CDF data.
Note that, in this figure (and in subsequent figures from the works of
Butensch\"on and Kniehl), it is clear that the color-singlet
contribution alone does not describe the data.

\begin{figure}[h!tb]
\begin{center}
\includegraphics[width=0.45\columnwidth]{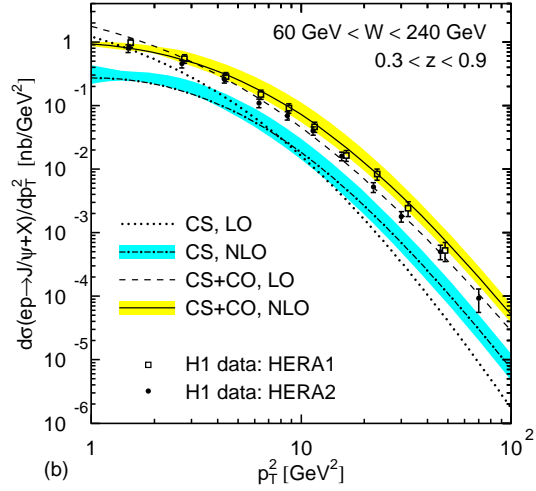}
\end{center}
\caption{The global fit of Butensch\"on and Kniehl
\cite{Butenschoen:2011yh} compared to the H1 data
\cite{Adloff:2002ex,Aaron:2010gz}. (Figure from
Ref.~\cite{Butenschoen:2011yh}.)
}
\label{buten-kniehl-H1}
\end{figure}

As can also be seen from Fig.~\ref{ma-buten-cdf}, the prediction of Ma, Wang,
and Chao deviates from the data at low $p_T$. Presumably, the prediction
of Butensch\"on and Kniehl would show a similar deviation if it were
extended to sufficiently low $p_T$. This deviation may be an indication
that factorization is breaking down, and/or that large logarithms of
$m_c^2/p_T^2$ must be be resummed in the low-$p_T$ region. As can be
seen in Fig.~\ref{buten-kniehl-H1}, 
there is also a slight discrepancy in shape in the fit of Butensch\"on
and Kniehl to the H1 data, which might be attributable to a breakdown of
factorization and/or a need to resum logarithms of $m_c^2/p_T^2$ in the
low-$p_T$ region.

\begin{figure}[h!tb]
\begin{center}
\includegraphics[width=0.45\columnwidth]{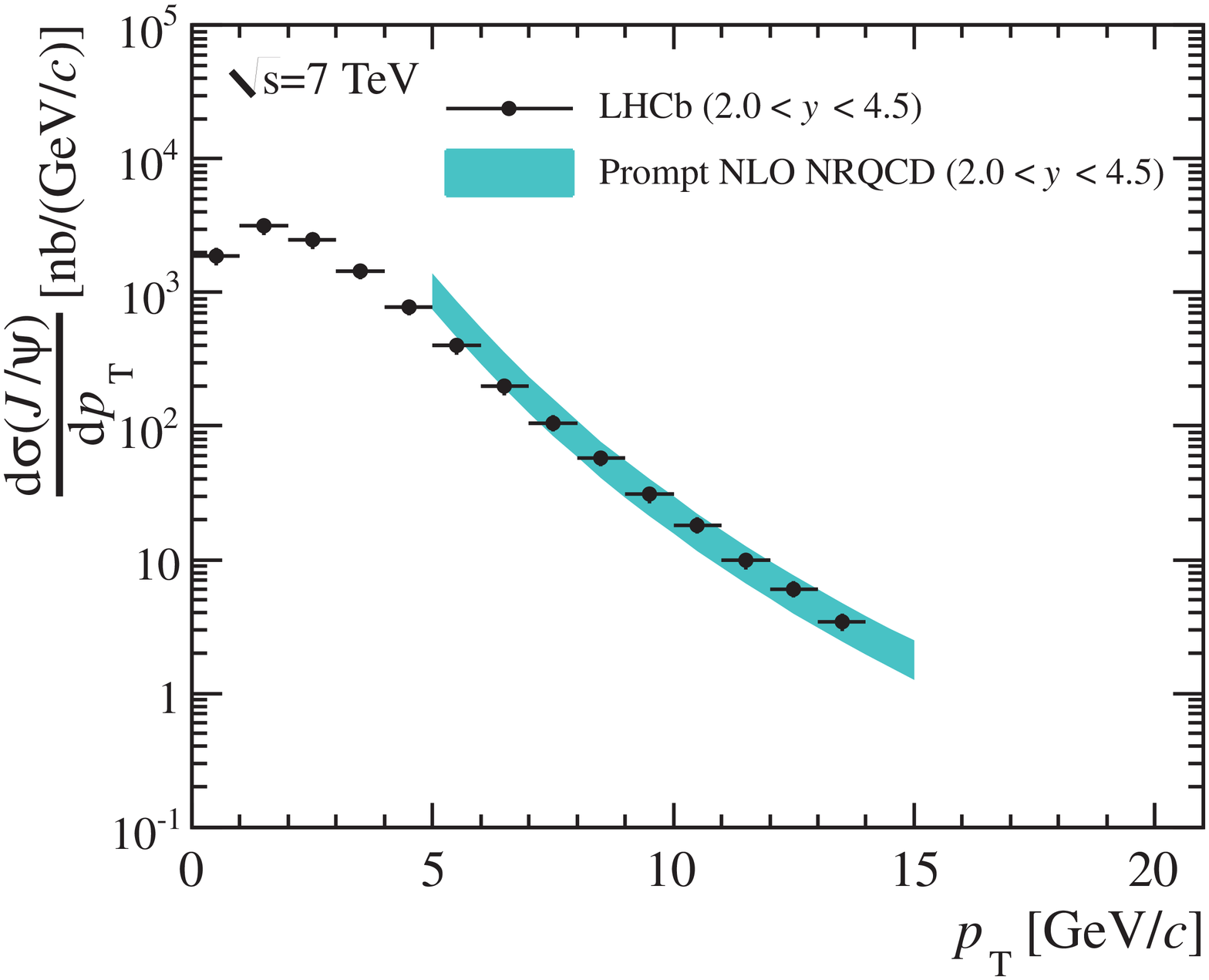}\qquad
\includegraphics[width=0.45\columnwidth]{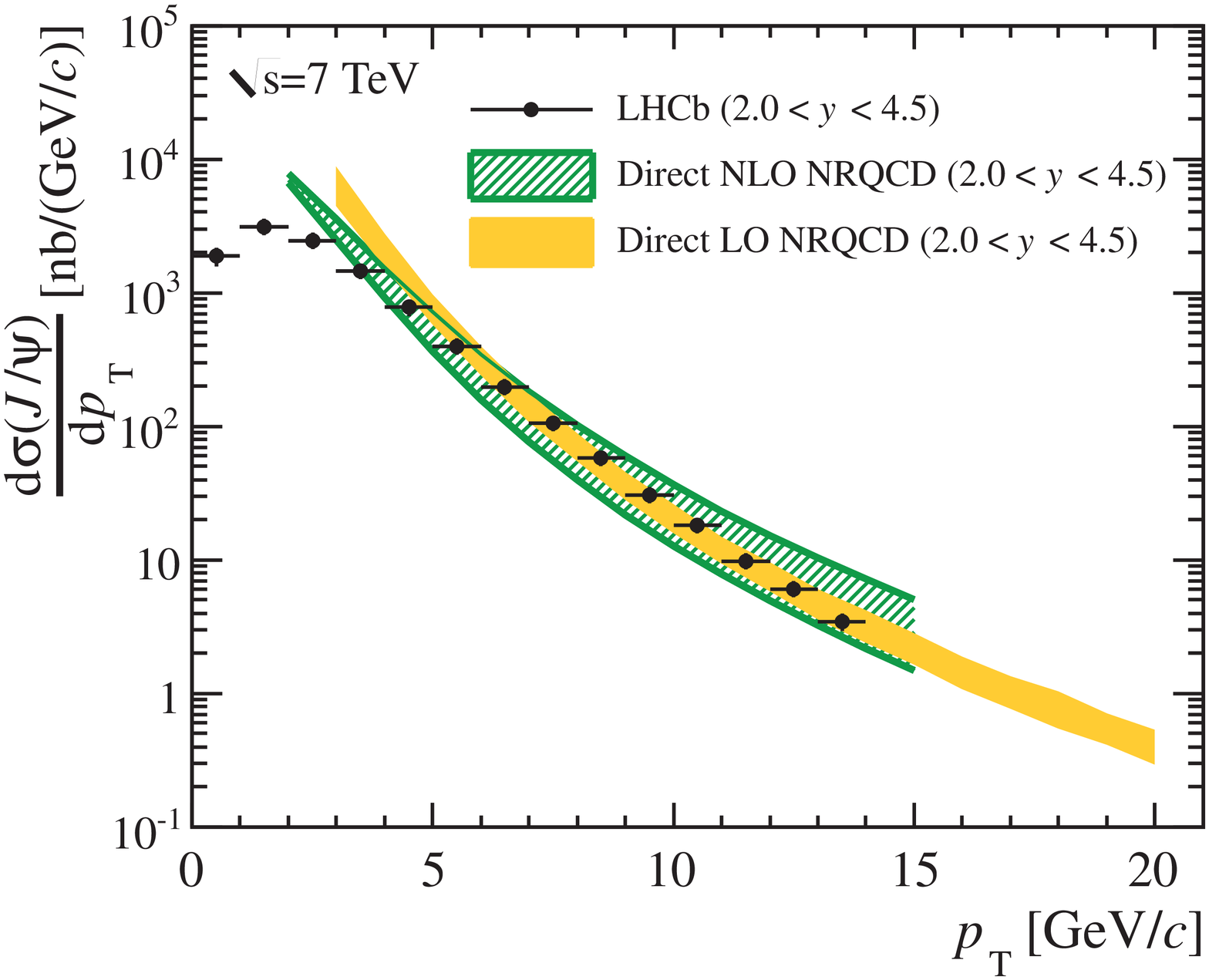}
\end{center}
\caption{Comparisons of NLO predictions with the LHCb
data \cite{Aaij:2011jh}. Left: The 
prediction of Ma, Wang, and Chao. (Figure from Ref.~\cite{Aaij:2011jh}.)
Right: The prediction of Butensch\"on and Kniehl from a fit 
to the Tevatron and HERA data \cite{Butenschoen:2010rq}. (Figure from 
Ref.~\cite{Aaij:2011jh}.)
\label{chao-buten-LHCb}}
\end{figure}

The fit of Ma, Wang, and Chao to the Tevatron data and the fit of
Butensch\"on and Kniehl \cite{Butenschoen:2010rq} to the Tevatron and
HERA data have been used to make predictions for $J/\psi$ cross sections
at the LHC. Comparisons to the LHCb data to the prediction of Ma, Wang,
and Chao and to the prediction of Butensch\"on and Kniehl are shown in
Fig.~\ref{chao-buten-LHCb}.

\begin{figure}[h!tb]
\begin{center}
\includegraphics[width=0.45\columnwidth]{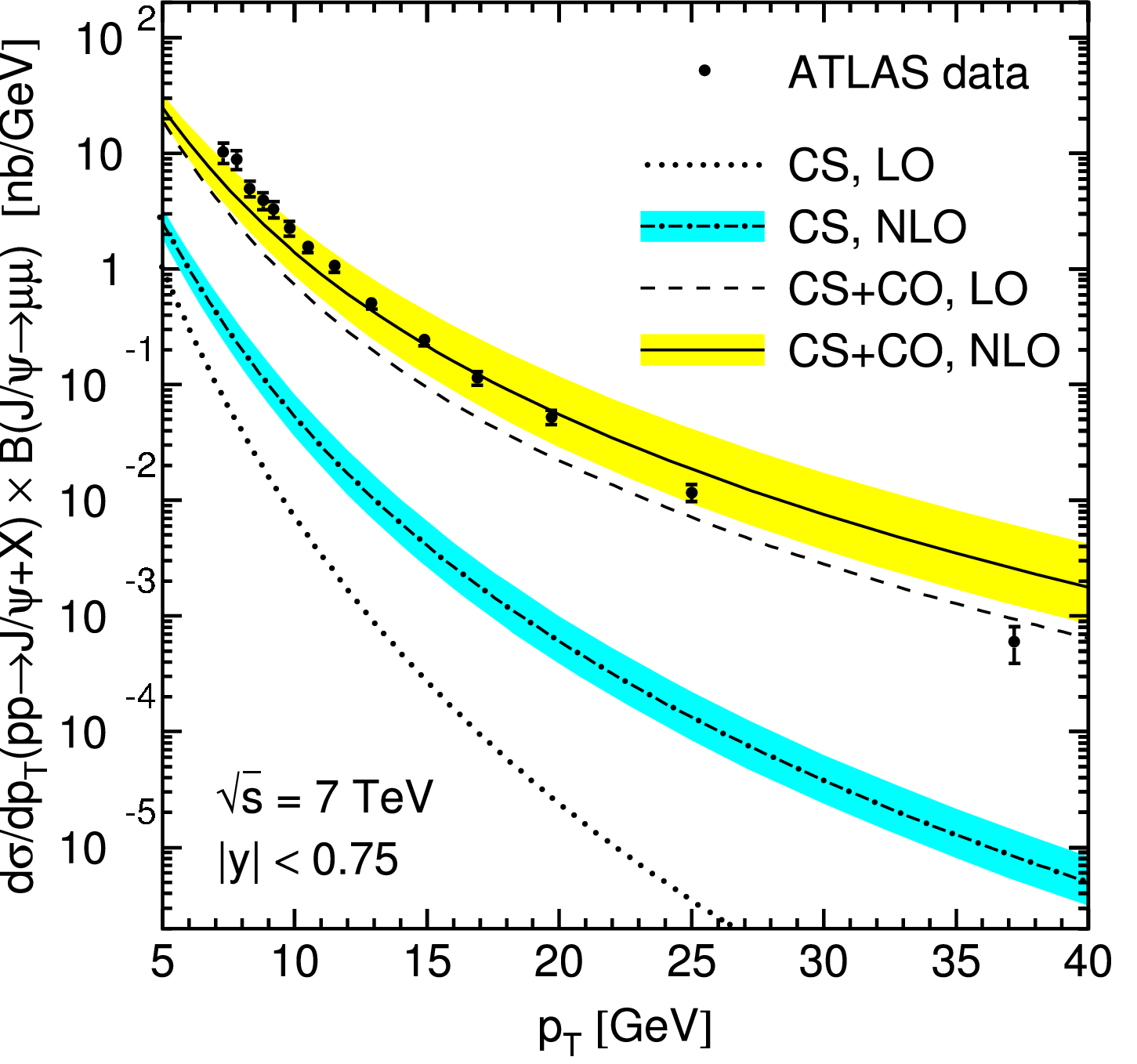}
\end{center}
\caption{Comparison of a prediction based on the global fit of
Butensch\"on and Kniehl \cite{Butenschoen:2011yh} with Atlas data
\cite{Aad:2011sp}. (Figure from Ref.~\cite{Butenschoen:2011yh}.)}
\label{BK-Atlas}
\end{figure}

As can be seen, the prediction of Ma, Wang, and Chao yields a
good fit to the data. The prediction of Butensch\"on and Kniehl also
fits the data within errors, but has the wrong shape: It overshoots the
data at high $p_T$. This shape discrepancy at high $p_T$ is even more
apparent in Fig.~\ref{BK-Atlas}, 
which shows a prediction of Butensch\"on and Kniehl, which is based of
their global fit, in comparison to Atlas data. Again, a resummation of
large logarithms of $p_T^2/m_c^2$ at high $p_T$ may improve the shape of
the Butensch\"on and Kniehl predictions, while worsening the shape of
the Ma, Wang, and Chao predictions.

\begin{figure}[h!tb]
\begin{center}
\includegraphics[width=0.45\columnwidth]{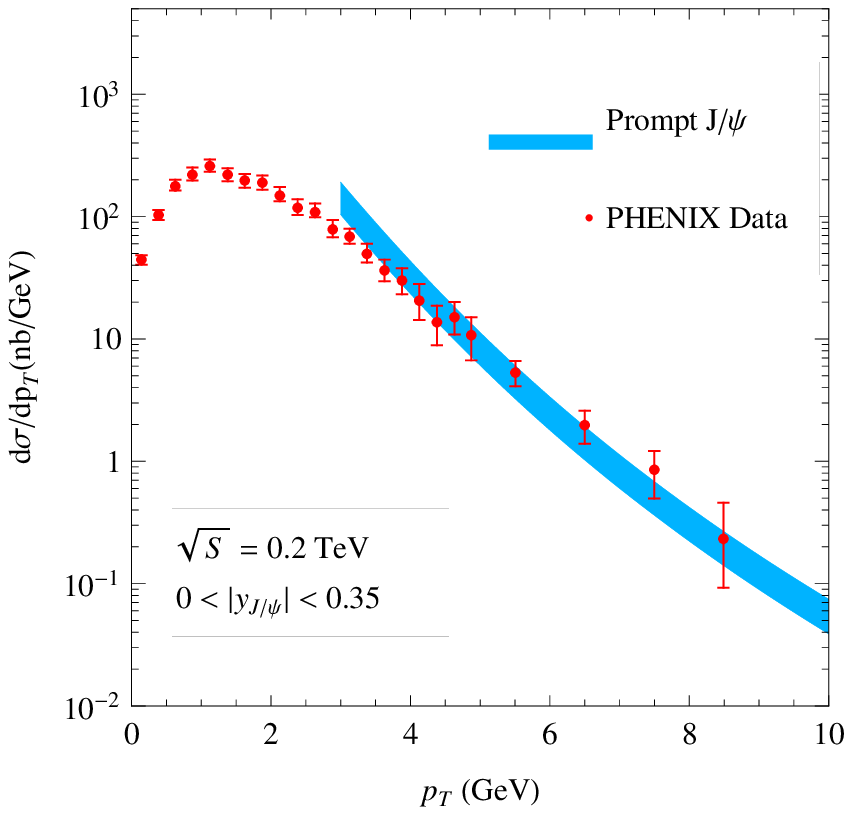}\quad
\includegraphics[width=0.45\columnwidth]{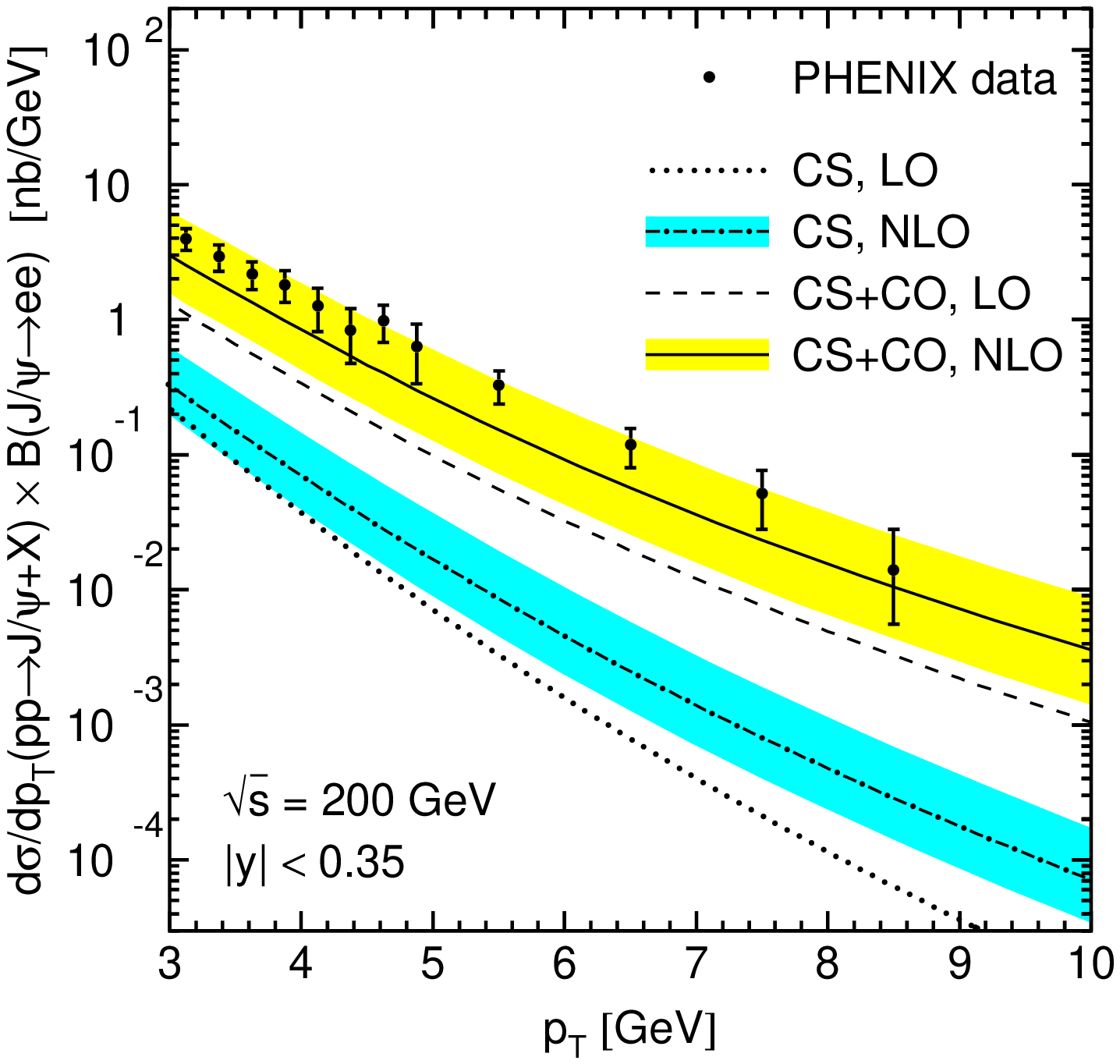}
\end{center}
\caption{Comparisons of NLO calculations with PHENIX data
\cite{Adare:2009js}. Left: The prediction of Ma, Wang, and Chao
\cite{Ma:2010jj}. (Figure from Ref.~\cite{Ma:2010jj}.) Right: The global 
fit of Butensch\"on and Kniehl \cite{Butenschoen:2010rq}. (Figure
from Ref.~\cite{Butenschoen:2010rq}.)
}
\label{chao-buten-PHENIX}
\end{figure}

Comparisons of PHENIX data with the prediction of Ma, Wang, and Chao and
the global fit of Butensch\"on and Kniehl are shown in
Fig.~\ref{chao-buten-PHENIX}. 
The prediction of Ma, Wang, and Chao agrees well with the data. This
prediction takes into account the effects of feeddown from the
$\chi_{cJ}$ states and the $\psi(2S)$ in both the fit to the CDF data
that is used to extract the LDMEs and in the prediction for the PHENIX
data. The global fit of Butensch\"on and Kniehl undershoots the PHENIX
data slightly, but still agrees within errors. Feeddown effects are not
taken into account in the global fit of Butensch\"on and Kniehl, but
they are expected to be small \cite{Butenschoen:2010rq}.

After this talk was presented, a new NLO calculation of $J/\psi$
hadroproduction by Gong, Wan, Wang, and Zhang \cite{Gong:2012ug}
appeared. This calculation includes contributions at NLO from feeddown
from the $\chi_{cJ}$ and $\psi(2S)$ states and extracts all three
color-octet NRQCD LDMEs from a combined fit to the CDF Run~II
\cite{Abulencia:2007us} and LHCb \cite{Aaij:2011jh} measurements of the
cross section, differential in $p_T$. The fit makes use only of data with 
$p_T\geq 7$~GeV. Gong {\it et al.}\ were able to
determine three linear combinations of color-octet
LDMEs with good accuracy. Their LDMEs yield
\begin{eqnarray}  
M_{0,r_0}&=&(6.0\pm 1.0)\times 10^{-2}~\hbox{GeV}^3,\nonumber\\
M_{1,r_1}&=&(0.07 \pm 0.01)\times 10^{-2}~\hbox{GeV}^3,           
\end{eqnarray} 
in good agreement with the fit of Ma, Wang, and Chao
[Eq.~(\ref{ma-wang-chao-fit})].

\subsection{Polarization}

Polarization of a ${}^3S_1$ quarkonium state is analyzed experimentally 
in terms of the angular distribution the lepton pairs into which it 
decays. The angular distribution is usually parametrized as
\begin{equation}
W(\theta,\phi)\propto 1+\lambda_\theta\cos^2\theta
+\lambda_\phi\sin^2\theta\cos(2\phi)
+\lambda_{\theta\phi}\sin(2\theta)\cos\phi,
\end{equation}
where $\theta$ and $\phi$ are the polar and azimuthal angles, 
respectively, of the positively charged lepton with respect to some 
coordinate system (often called the ``frame'') and $\lambda_\theta$, 
$\lambda_\phi$, and $\lambda_{\theta\phi}$ are the measured polarization 
parameters. The parameter $\lambda_\phi$, which is also called $\alpha$ 
in the literature, is a measure of the amount of transverse or 
longitudinal polarization of the quarkonium: $\lambda_\phi=1$ 
corresponds to completely transverse polarization, and $\lambda_\phi=-1$ 
corresponds to completely longitudinal polarization. 

Quarkonium polarization at large $p_T$ has long been thought to provide 
a definitive test of the color-octet production mechanism. However, this 
is an idea that is based on arguments at LO in $\alpha_s$, and, as we will 
see, it may not be correct at NLO.

\subsection{Polarization at LO in $\bm{\alpha_s}$}

At LO in $\alpha_s$, hadroproduction of a ${}^3S_1$ quarkonium at large
$p_T$ is dominated by the process of gluon fragmentation into a
color-octet ${}^3S_1$ $Q\bar Q$ pair.  At large $p_T$ in the hadronic
center-of-momentum (CM) frame, the gluon is nearly on mass shell, and,
so, is transversely polarized. At LO in $\alpha_s$, that transverse
polarization is transferred completely to the $Q\bar Q$ pair in the
perturbative fragmentation process. As the $Q\bar Q$ pair evolves
nonperturbatively into a ${}^3S_1$ quarkonium through the emission of
gluons that are soft in the $Q\bar Q$ rest frame, the transverse
polarization is preserved at LO in $v$ because of the heavy-quark spin
symmetry \cite{Cho:1994ih}.\footnote{There are spin-flip interactions,
which are suppressed as $v^3$. This suppression has been  verified in a
lattice calculation of {\it decay} matrix elements
\cite{Bodwin:2005gg}.} The transverse polarization is defined in the
hadronic CM frame relative to the direction of the fragmenting gluon's
momentum, and is invariant under a boost in that direction. At LO in
$\alpha_s$ and $v$, that boost vector coincides with the boost vector
from the quarkonium rest frame to the hadronic CM frame, which defines
the polar axis of the ``helicity frame.'' Hence, at LO in $\alpha_s$
and $v$, one expects $\lambda_\theta$ in the helicity frame to approach
$1$ as $p_T$ increases.

\begin{figure}[h!tb]
\begin{center}
\includegraphics[width=0.45\columnwidth]{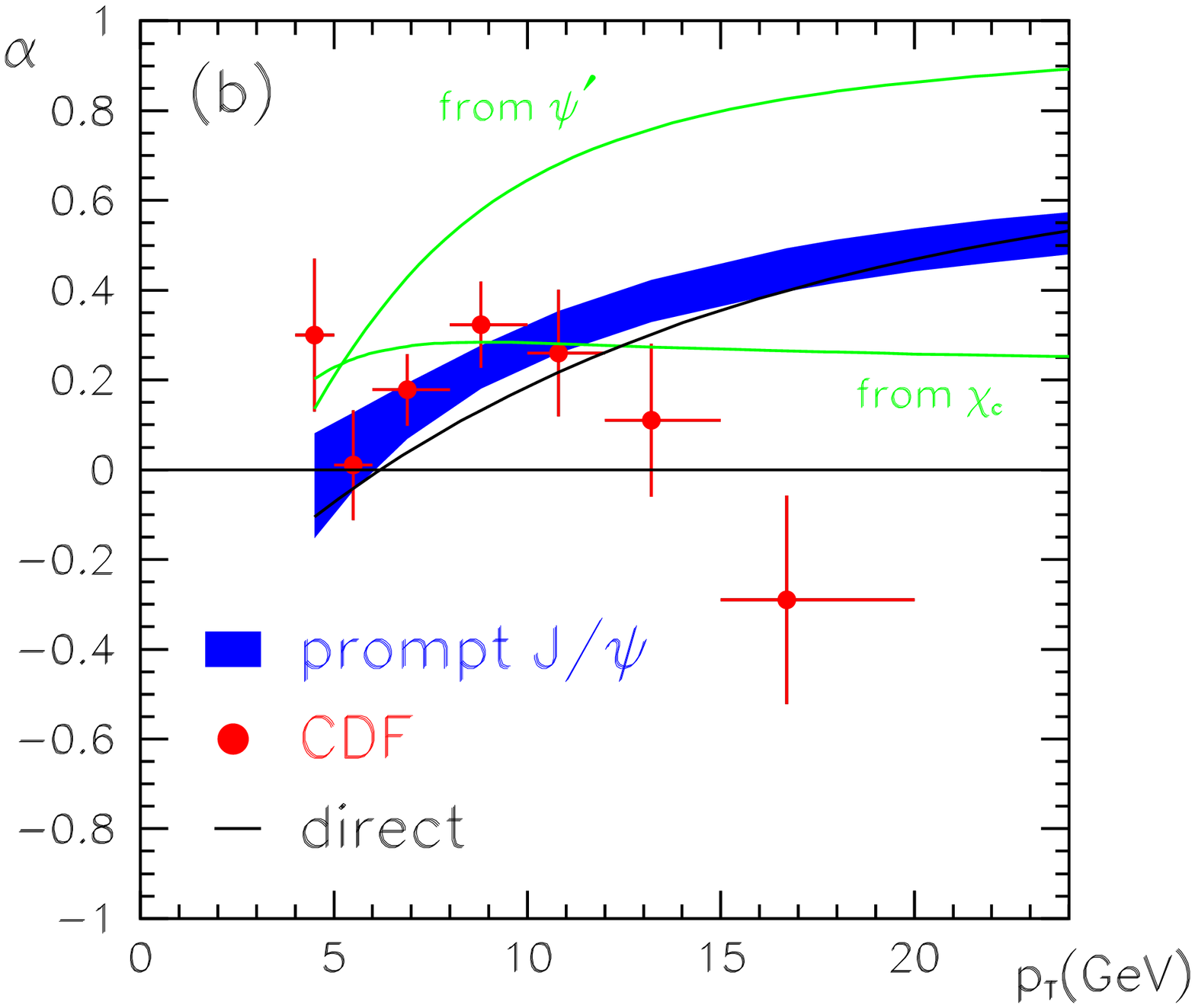}\quad
\includegraphics[width=0.45\columnwidth]{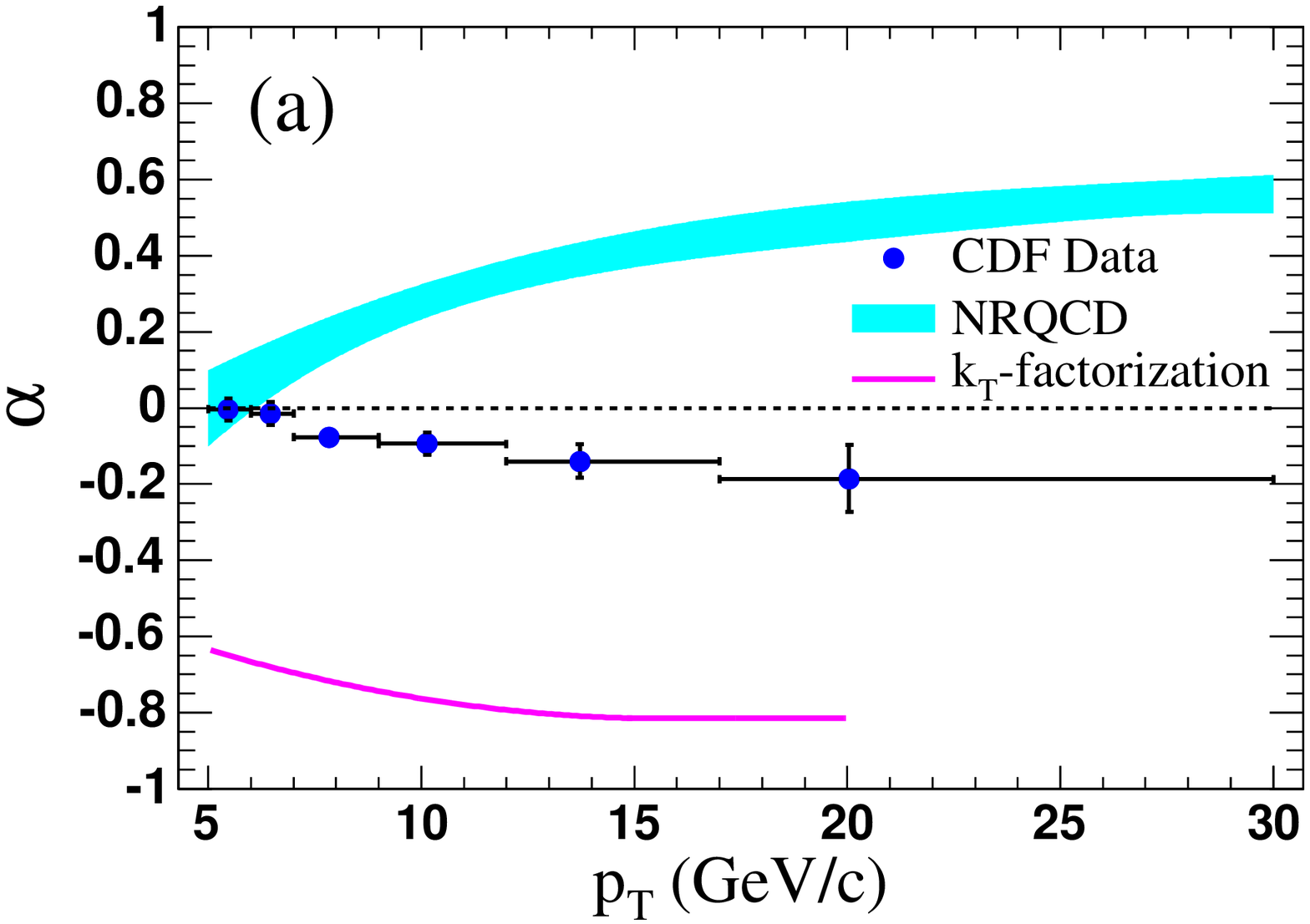}
\end{center}
\caption{Comparisons of the LO NRQCD prediction for polarization of the 
$J/\psi$ \cite{Braaten:1999qk} with CDF data. Left: The CDF Run~I data 
\cite{Affolder:2000nn}. (Figure provided by Jungil Lee.) Right: 
The CDF Run~II data \cite{Abulencia:2007us}. (Figure from 
Ref.~\cite{Abulencia:2007us}.)
}
\label{cdf-pol}
\end{figure}

The LO prediction of transverse polarization at large $p_T$ has not been
confirmed experimentally. As can be seen from the left-hand plot in
Fig.~\ref{cdf-pol}, 
the LO prediction agrees, within errors, with the
CDF Run~I measurement, except for the two highest $p_T$ points.
However, as can be seen from right-hand plot in Fig.~\ref{cdf-pol}, the
LO prediction disagrees completely with the CDF Run~II measurement.
Feeddown contributions from the $\chi_c$ states (about $30\%$) and the
$\psi(2S)$ state (about 10\%) are included in both the data and the
theoretical prediction. The CDF Run~I and Run~II measurements are also
in disagreement with each other. The reason for this discrepancy has not
been found.

\begin{figure}[h!tb]
\begin{center}
\includegraphics[width=0.45\columnwidth]{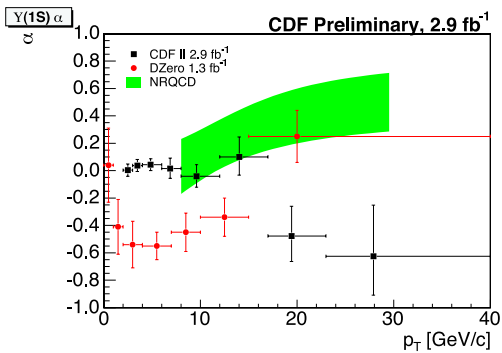}
\end{center}
\caption{Comparison of the LO NRQCD prediction for polarization of the
$\Upsilon$ \cite{Braaten:1999qk} with the CDF Run~II data
\cite{CDF-note-9966} and the D0 Run~II data \cite{Abazov:2008aa}.
(Figure provided by Hee Sok Chung, using
Refs.~\cite{CDF-note-9966,Abazov:2008aa}.)}
\label{upsilon-pol}
\end{figure}

The LO NRQCD factorization prediction for polarization of the
$\Upsilon(1S)$ is compared with the CDF and D0 Run~II data in
Fig.~\ref{upsilon-pol}. 
Both the CDF and D0 results are incompatible
with the LO prediction, but in different regions of $p_T$. The CDF and
D0 results are also incompatible with each other. No reason for the
incompatibility of the experimental results has been found.

\subsubsection{First Complete NLO Calculations of Polarization}

The first NLO calculation of $J/\psi$ polarization in 
photoproduction was carried out by Butensch\"on and Kniehl 
\cite{Butenschoen:2011ks}. This calculation is complete in the sense 
that it includes all of the production channels that appear in NRQCD 
through relative order $v^4$ [Eq.~(\ref{S-wave-MEs})]. It was preceded 
by a calculation at NLO for the ${}^3S_1$ color-singlet channel for 
$J/\psi$ photoproduction. \cite{Artoisenet:2009xh}.

The first complete NLO calculations of $J/\psi$ polarization in
hadroproduction were carried out by Butensch\"on and Kniehl
\cite{Butenschoen:2012px} and by Chao, Ma, Shao, Wang and Zhang
\cite{Chao:2012iv}. These calculations also include all of the
production channels that appear in NRQCD through relative order $v^4$.
They were preceded by partial calculations at NLO for the ${}^3S_1$
color-singlet channel for $J/\psi$ hadroproduction \cite{Gong:2008sn}
and $\Upsilon(1S)$ hadroproduction \cite{Artoisenet:2008fc} and by
calculations at NLO for the ${}^3S_1$ and ${}^1S_0$ color-octet channels
for $J/\psi$ hadroproduction \cite{Gong:2008ft} and $\Upsilon(1S)$
hadroproduction \cite{Gong:2010bk}.

At NLO, the theoretical predictions for polarization change
dramatically. Polarization in the color-singlet channel changes from
transverse to longitudinal \cite{Artoisenet:2008fc,Gong:2008sn}.
However, this channel makes a very small contribution to the
hadroproduction cross section. A much more important issue is that, at
NLO at high $p_T$, there are large corrections to production in the
${}^3P_J$ color-octet channel that are mostly transversely polarized. At
NLO, the ${}^3S_1$ color-octet channel is still transversely polarized
at high $p_T$. However, owing to the large NLO corrections in the
${}^3P_J$ color-octet channel, it is no longer the dominant channel for
hadroproduction at high $p_T$, and it is not the only source of
transverse polarization.

\begin{figure}[h!tb]
\begin{center}
\includegraphics[width=0.45\columnwidth]{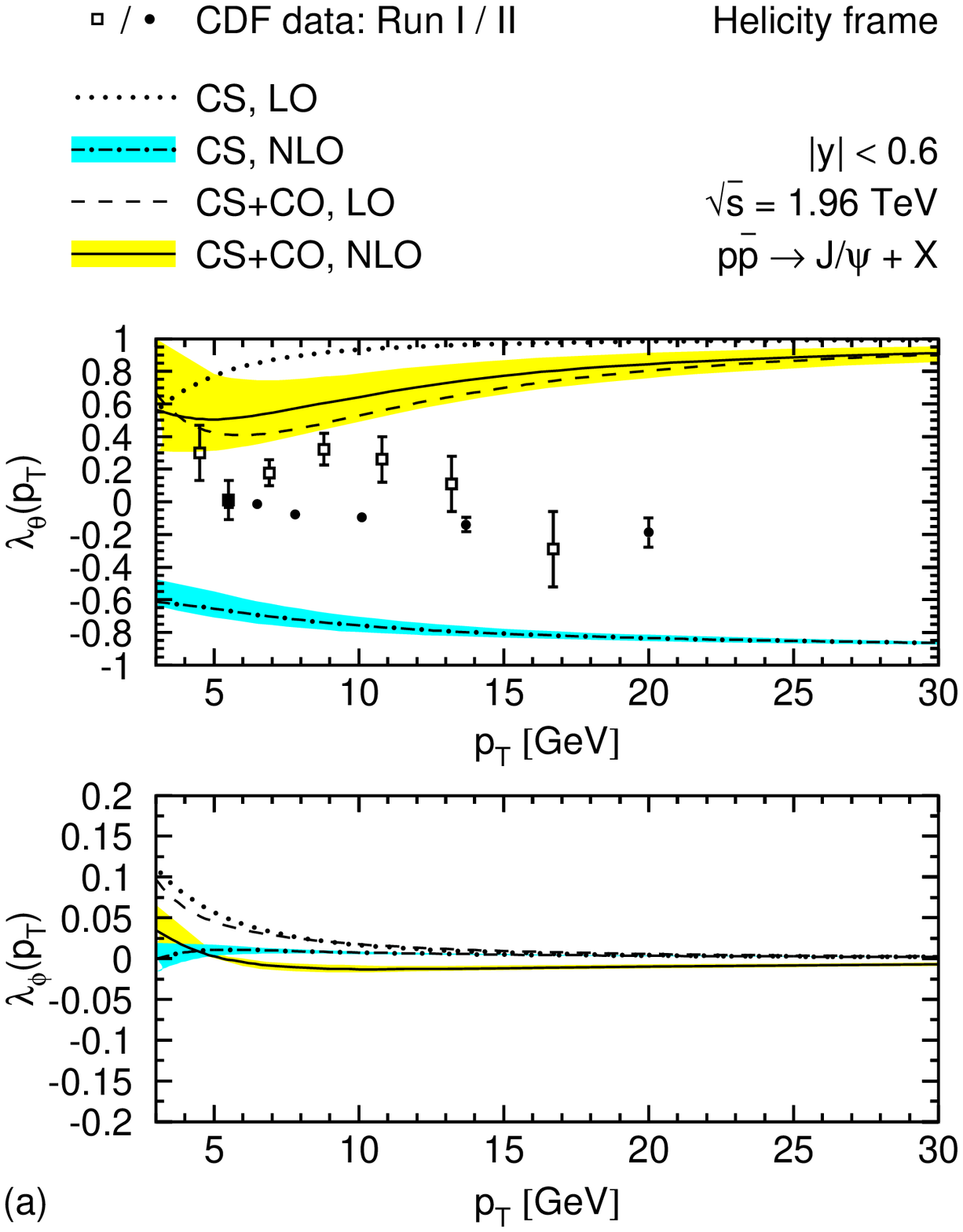}\qquad
\includegraphics[width=0.45\columnwidth]{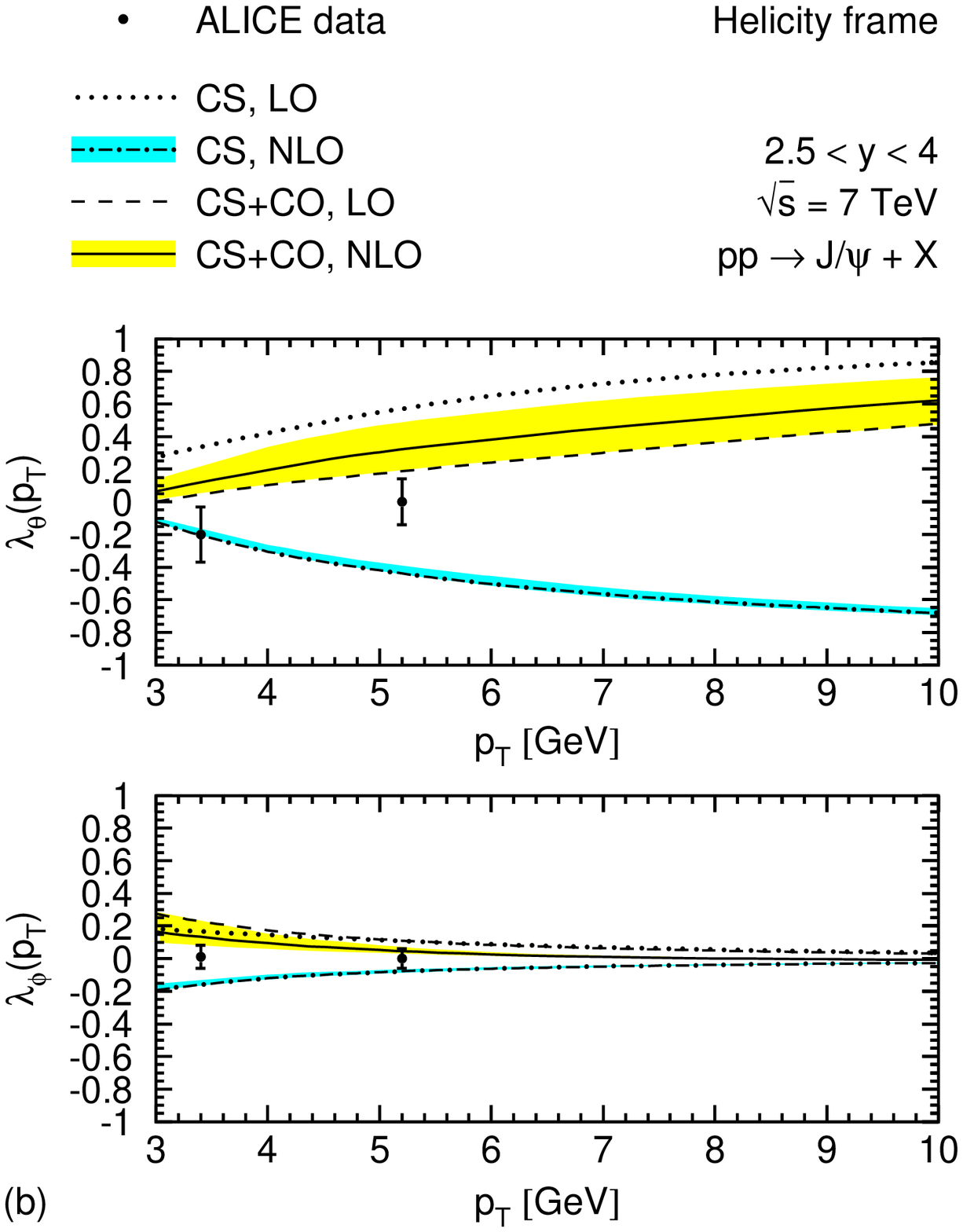}
\end{center}
\caption{Comparisons of the NLO NRQCD prediction of Butensch\"on and
Kniehl \cite{Butenschoen:2012px} for the polarization of the $J/\psi$
with hadroproduction data. Left: the CDF Run~I \cite{Affolder:2000nn}
and Run~II \cite{Abulencia:2007us} data. (Figure from
Ref.~\cite{Butenschoen:2012px}.) Right: Alice data \cite{Abelev:2011md}.
(Figure from Ref.~\cite{Butenschoen:2012px}.)}
\label{BK-pol}
\end{figure}

Fig.~\ref{BK-pol} 
shows the prediction of Butensch\"on and
Kniehl for $J/\psi$ polarization in hadroproduction in comparison with 
the CDF Run~I and Run~II data and Alice data.
The prediction is based on the LDMEs from the global fit of Butensch\"on
and Kniehl \cite{Butenschoen:2011yh}. It is in considerable disagreement
with the data. On the other hand, as can be seen from
Fig.~\ref{BK-pol}, the NLO $J/\psi$ polarization prediction of
Butensch\"on and Kniehl is in rough agreement with the Alice data. It
should be noted, though, that the Alice data include $J/\psi$ production
from $B$-meson decays and feeddown from the $\chi_{cJ}$ and $\psi(2S)$
states.

\begin{figure}[h!tb]
\begin{center}
\includegraphics[width=0.45\columnwidth]{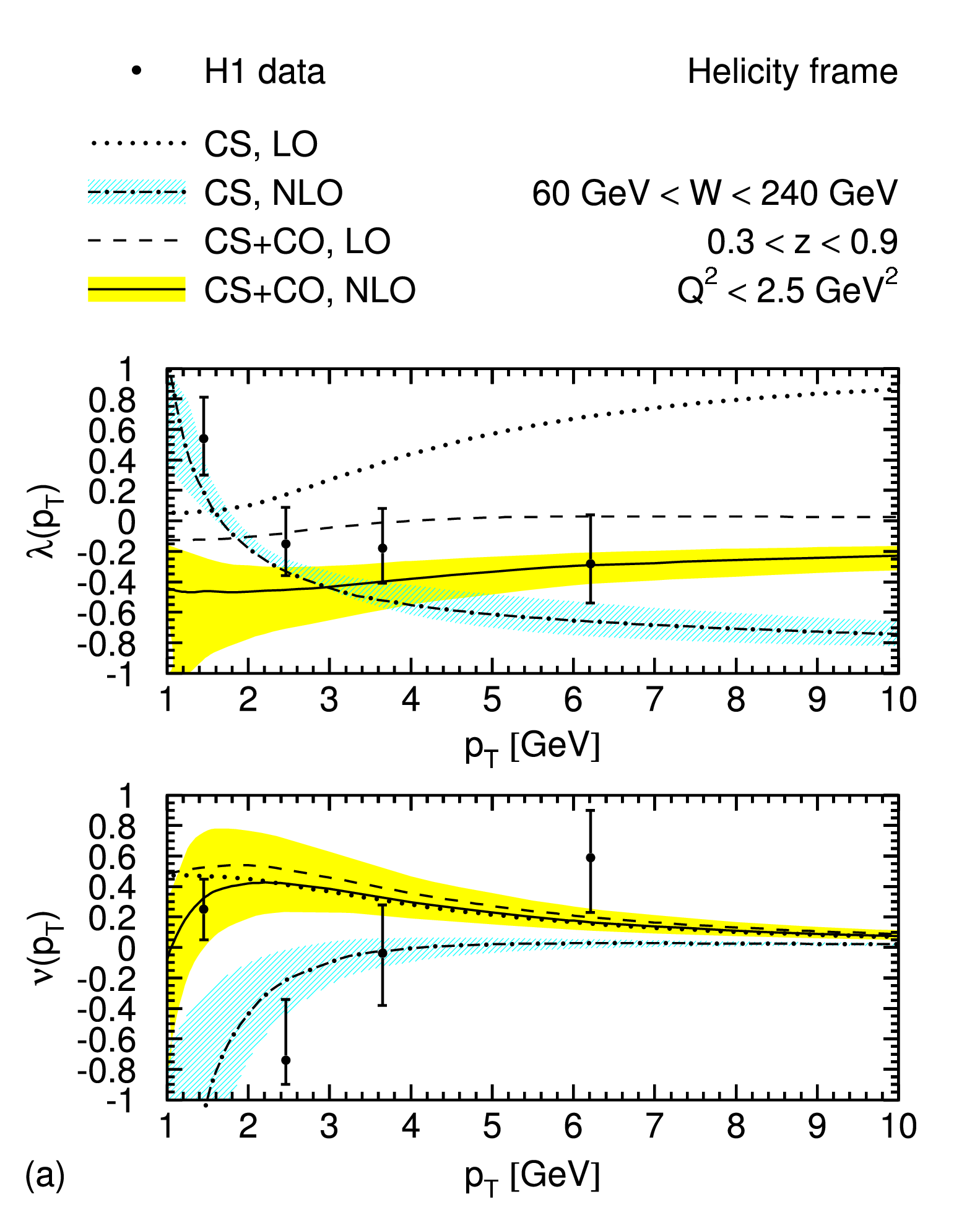}
\end{center}
\caption{Comparison of the NLO NRQCD prediction of Butensch\"on and 
Kniehl \cite{Butenschoen:2012px} for the polarization of the
$J/\psi$ with H1 data \cite{Aaron:2010gz,Chekanov:2009ad}. (Figure from 
Ref.~\cite{Butenschoen:2012px}.)}
\label{BK-H1-pol}
\end{figure}

The LDMEs from the global fit of Butensch\"on and Kniehl can also be
used to predict the polarization of the $J/\psi$ in photoproduction at
HERA. That prediction is compared in Fig.~\ref{BK-H1-pol} 
with H1 data. The data are roughly compatible with the theory at large
$p_T$, but the error bars are large.

In contrast, Chao, Ma, Shao, Wang, and Zhang \cite{Chao:2012iv}
determined the three color-octet NRQCD LDMEs by using data from the CDF
Run~II measurements of the $J/\psi$ cross section, differential in
$p_T$, and the CDF Run~I and Run~II measurements of the $J/\psi$
polarization, differential in $p_T$. The resulting LDMEs are barely
compatible with the LDMEs from the fit of Ma, Wang, and Chao
\cite{Ma:2010yw,Ma:2010jj} to the production cross section alone: The
values of $M_{0,r_0}$ and $M_{1,r_1}$ that Chao {\it et al.}\ obtain
differ from those of Ma, Wang, and Chao by about $2\sigma$. This tension
between the two sets of LDMEs suggests that the LDMEs from the fit of
Ma, Wang, and Chao might lead to a polarization prediction that deviates
from the CDF Run~II data, although perhaps with large uncertainties.
 
\begin{figure}[h!tb]
\begin{center}
\includegraphics[width=0.45\columnwidth]{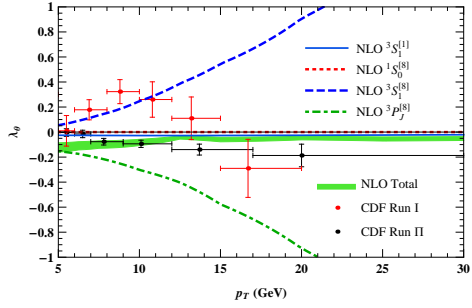}
\end{center}
\caption{Comparison of the NLO NRQCD prediction of 
Chao, Ma, Shao, Wang, and Zhang \cite{Chao:2012iv} for the polarization
of the $J/\psi$ with the CDF Run~I \cite{Affolder:2000nn} and Run~II
\cite{Abulencia:2007us} data. (Figure from Ref.~\cite{Chao:2012iv}.)}
\label{Chao-CDF-pol}
\end{figure}

\begin{figure}[h!tb]
\begin{center}
\includegraphics[width=0.45\columnwidth]{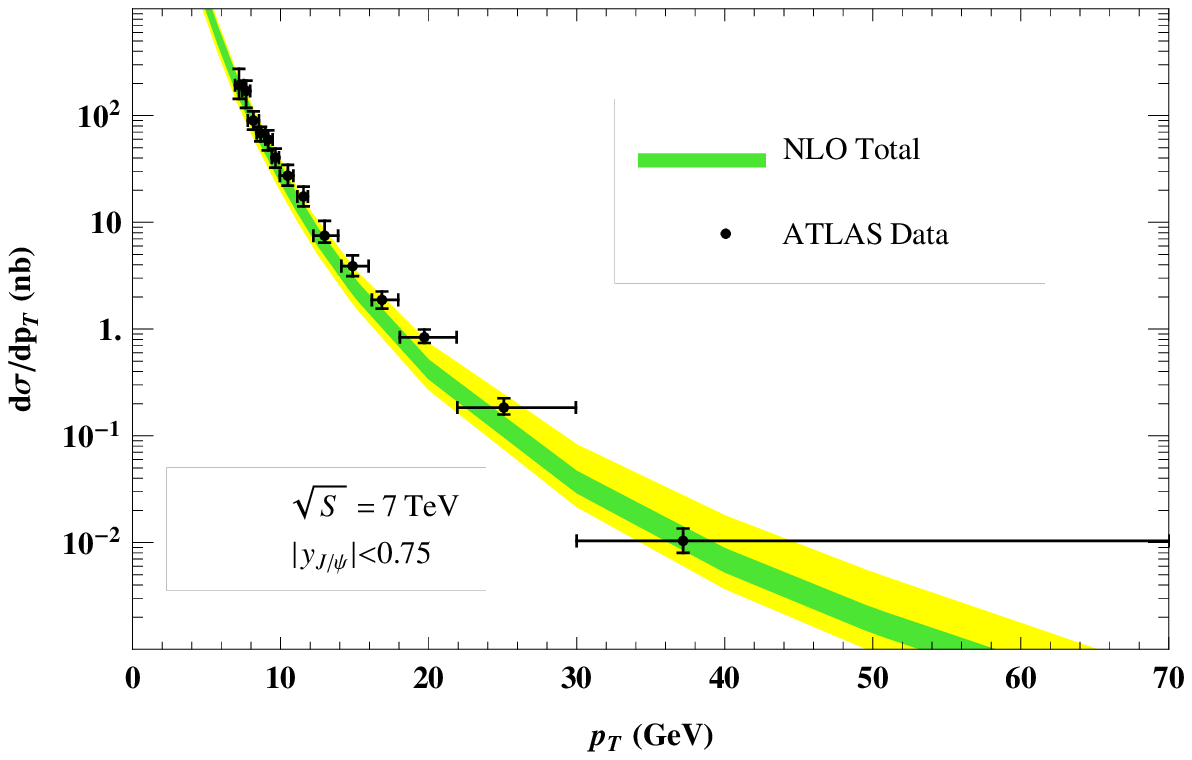}\quad
\includegraphics[width=0.45\columnwidth]{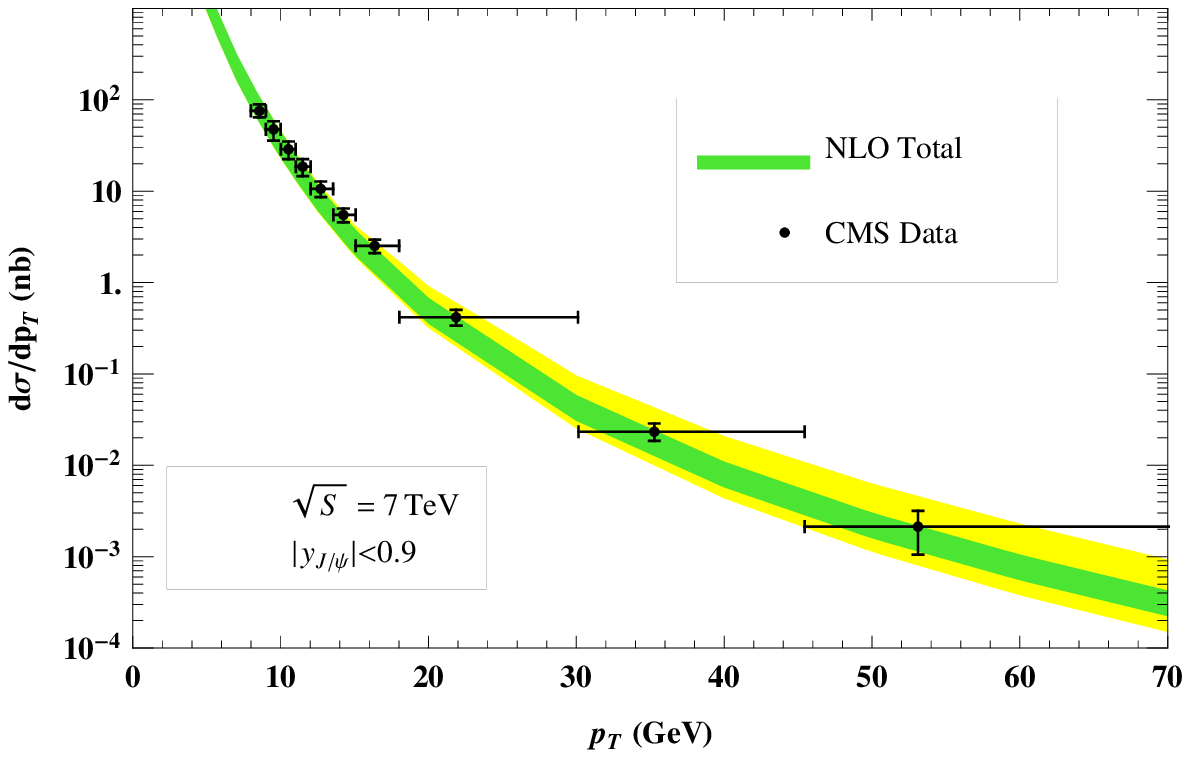}
\end{center}
\caption{Comparisons of LHC data with the NLO NRQCD prediction for the
$J/\psi$ cross section of Chao, Ma, Shao, Wang, and Zhang
\cite{Chao:2012iv}. The prediction uses NRQCD LDMEs that are extracted
from fits to the CDF measurements of the $J/\psi$ cross section
\cite{Acosta:2004yw} and the $J/\psi$ polarization
\cite{Abulencia:2007us,Affolder:2000nn}. Left: The Atlas data
\cite{Aad:2011sp}. Right: The CMS data \cite{Chatrchyan:2011kc}. (Figure
from Ref.~\cite{Chao:2012iv}.)
}
\label{Chao-LHC-prod}
\end{figure}

\begin{figure}[h!tb]
\begin{center}
\includegraphics[width=0.45\columnwidth]{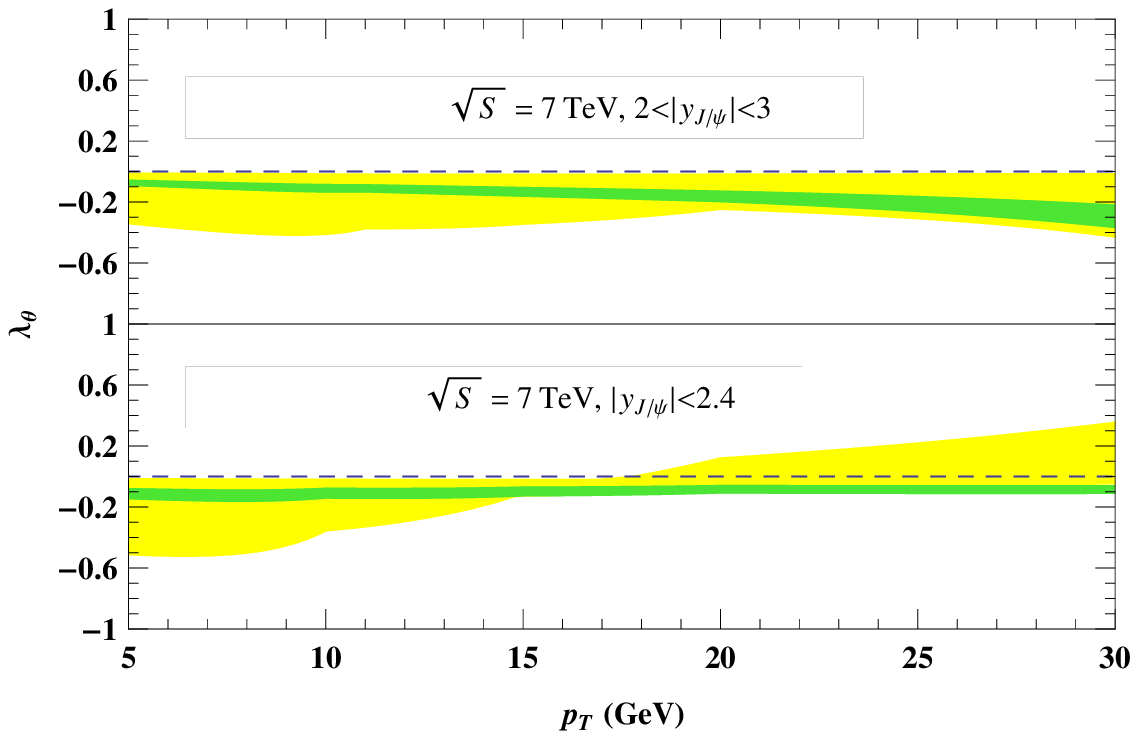}
\end{center}
\caption{NLO NRQCD prediction of 
Chao, Ma, Shao, Wang, and Zhang \cite{Chao:2012iv}
for the polarization of the $J/\psi$ at the LHC.
(Figure from Ref.~\cite{Chao:2012iv}.)}
\label{Chao-LHC-pol}
\end{figure}

\begin{figure}[h!tb]
\begin{center}
\includegraphics[width=0.45\columnwidth]{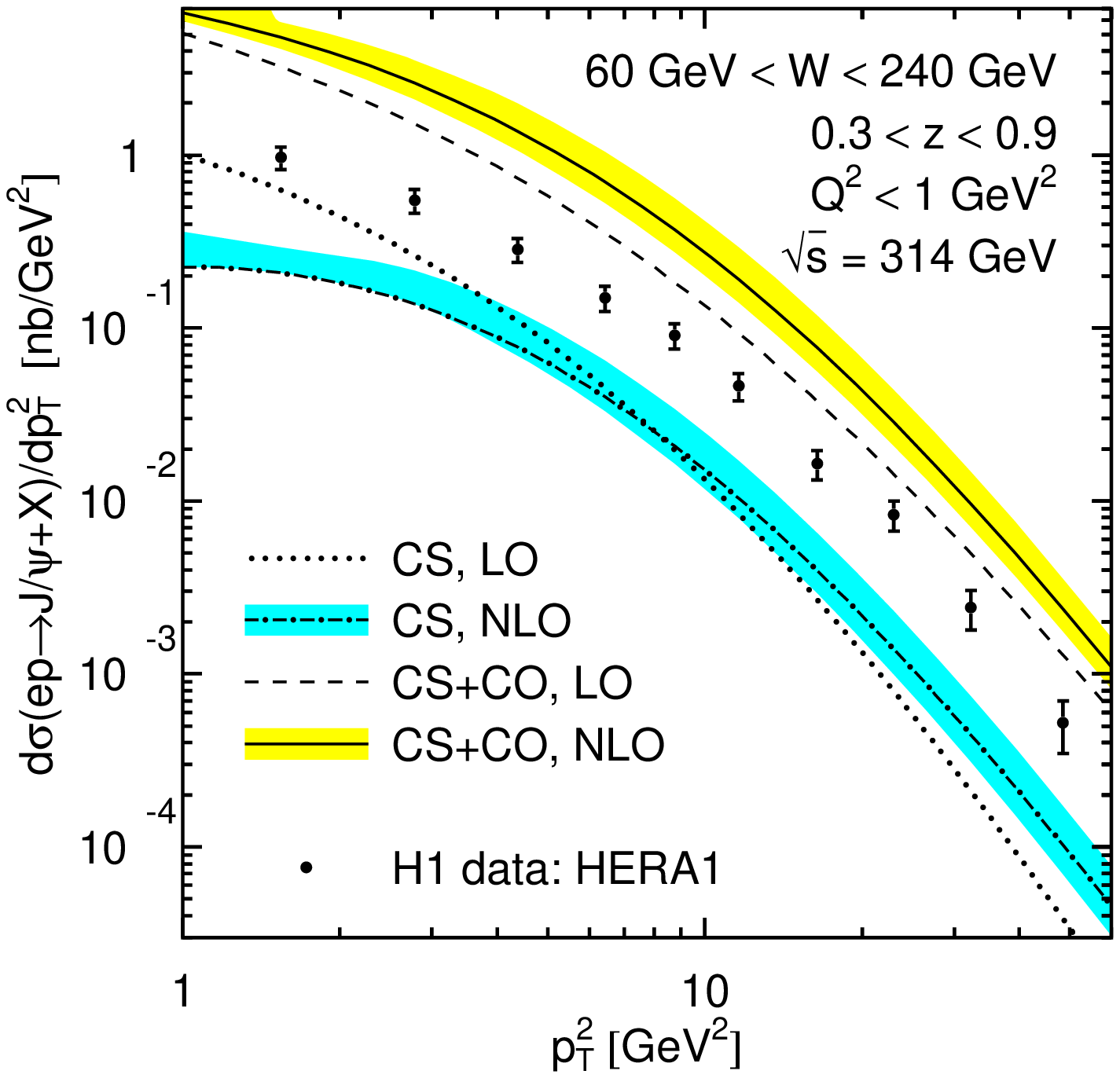}
\end{center}
\caption{Comparison of a prediction based on the LDMEs of Chao, Ma, 
Shao, Wang, and Zhang \cite{Chao:2012iv} with the 
H1 data \cite{Adloff:2002ex,Aaron:2010gz}. (Figure provided by 
Mathias Butensch\"on.)}
\label{Chao-Buten-H1}
\end{figure}

The LDMEs of Chao {\it et al.}\ lead to the prediction for $J/\psi$
polarization at CDF that is shown in Fig.~\ref{Chao-CDF-pol}. 
With this choice of LDMEs, the contributions of the color-octet
${}^3S_1$ and ${}^3P_J$ channels to the transverse polarization largely
cancel, leaving a net polarization parameter $\lambda_\theta$ that is
near zero, in agreement with the CDF Run~II data. This choice of  LDMEs
still gives reasonable predictions for the $J/\psi$ production cross
sections that have been measured by the Atlas and CMS Collaborations, as
can be seen from Fig.~\ref{Chao-LHC-prod}.
This reflects the fact that the $J/\psi$ LDMEs and, hence, the $J/\psi$
polarization in hadroproduction, are not well determined from fits to
the production cross section alone. The Chao {\it et al.}\ LDMEs lead to
a prediction of a slight longitudinal polarization at the LHC, as can be
seen in Fig.~\ref{Chao-LHC-pol}. 
However, in Fig.~\ref{Chao-Buten-H1}, 
one can see that the Chao {\it et al.}\ LDMEs seem to be incompatible
with the HERA photoproduction data, even at large $p_T$.

After this talk was presented, Gong {et al.}\ also completed a new NLO
calculation of $J/\psi$ polarization in hadroproduction
\cite{Gong:2012ug}, including the effects at NLO of the feeddown
contributions from the $\chi_{cJ}$ and $\psi(2S)$ states. Using the
color-octet LDMEs from their fits to the CDF Run~II and LHCb data, Gong
{et al.}\ made predictions for the $J/\psi$ polarization. Their
prediction for the polarization at CDF is intermediate between that of
Butensch\"on and Kniehl \cite{Butenschoen:2012px} and that of Chao
{\it et al.}\ \cite{Chao:2012iv}, agreeing with the CDF Run~I data
\cite{Affolder:2000nn}, except for the highest-$p_T$ point, but
disagreeing with the CDF Run~II data \cite{Abulencia:2007us}. Since the
polarization predictions of Gong {et al.}\ rely only on data with
$p_T\geq 7$~GeV, they are less subject to questions about the validity
of factorization than the predictions of Butensch\"on and Kniehl 
\cite{Butenschoen:2012px}.

\subsection{$\bm{e^+e^-\to J/\psi+X(\hbox{non-$c\bar c$})}$}

The total cross section for the process $e^+e^-\to 
J/\psi+X(\hbox{non-$c\bar c$})$ has been measured by the Belle 
Collaboration \cite{Pakhlov:2009nj}. 
Their result is
\begin{equation}                                             
\sigma(e^+e^-\rightarrow                                     
J/\psi+X(\hbox{non-}c\bar c)) =0.43\pm 0.09\pm 0.09\hbox{~pb}.
\end{equation}
NLO calculations using the NRQCD factorization approach have been
carried  out by Zhang, Ma, Wang, and Chao \cite{Zhang:2009ym} and by
Butensch\"on and Kniehl \cite{Butenschoen:2011yh} and lead to the 
prediction
\begin{equation}
\sigma(e^+e^-\rightarrow J/\psi+X(\hbox{non-}c\bar
c))=0.99_{-0.17}^{+0.35}\hbox{ pb}\qquad (\mu=\sqrt{s}/2),
\end{equation}
which is based on the NRQCD LDMEs from the global fit of Butensch\"on
and Kniehl \cite{Butenschoen:2011yh} and includes the effects of
feeddown from the $\chi_{cJ}$ and $\psi(2S)$ states \cite{Zhang:2009ym}.
The comparison of the theoretical calculations with the measured cross
section favors the value of $M_{0,r_0}$ from the global fit of
Butensch\"on and Kniehl, rather than the value from the fit of Ma, Wang,
and Chao \cite{Ma:2010yw,Ma:2010jj}.

There is tension between the Belle data and the theoretical prediction. 
However, it is worth noting that the most recent Belle measurements 
\cite{Pakhlov:2009nj} imply that
\begin{eqnarray}
\sigma(e^+e^-\rightarrow J/\psi +X)&=&\sigma(e^+e^-\rightarrow J/\psi+
c\bar c+X)+\sigma(e^+e^-\rightarrow J/\psi+X(\hbox{non-}c\bar c))
\nonumber\\
&=& 1.17\pm 0.12^{+0.13}_{-0.12}~\hbox{pb},
\end{eqnarray}
while the BaBar Collaboration obtained \cite{Aubert:2001pd}
\begin{equation}
\sigma(e^+e^-\rightarrow J/\psi +X)=2.52\pm 0.21\pm 0.21~\hbox{pb}.
\end{equation}
This suggests the possibility that the Belle value of
$\sigma(e^+e^-\rightarrow  J/\psi+X(\hbox{non-}c\bar c))$ is too small.
Furthermore, most of the Belle data are at $p_T<3$~GeV. Even if NRQCD
factorization could be established at high $p_T$, it might not hold at
such small values of $p_T$.

\subsection{$\bm{J/\psi}$ Production in $\bm{\gamma\gamma}$ Scattering at LEP~II}

\begin{figure}[h!tb]
\begin{center}
\includegraphics[width=0.45\columnwidth]{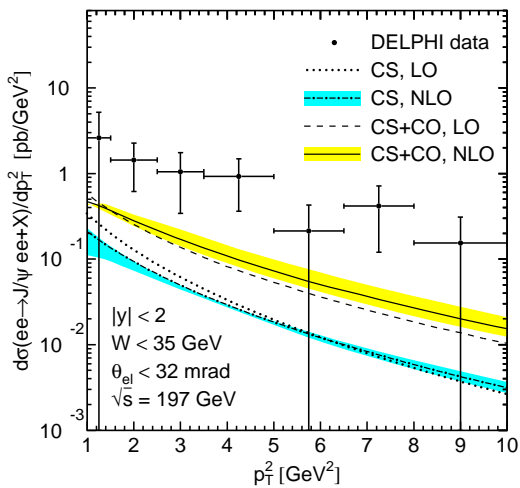}
\end{center}
\caption{Comparison of the global fit of Butensch\"on and Kniehl
\cite{Butenschoen:2011yh} with the DELPHI LEP~II data
\cite{Abdallah:2003du}. (Figure from Ref.~\cite{Butenschoen:2011yh}.)}
\label{buten-kniehl-DELPHI}
\end{figure}

Butensch\"on and Kniehl have also made an NLO calculation of $J/\psi$
production in $\gamma\gamma$ collisions at LEPII and incorporated it
into their global fit \cite{Butenschoen:2011yh}. This global fit is
compared with DELPHI data in Fig.~\ref{buten-kniehl-DELPHI}.
The DELPHI data are slightly incompatible with the global fit. However, 
the error bars are large, especially at large $p_T$, and one should keep 
in mind that NRQCD factorization, even if it is established at high 
$p_T$,  may not hold at such low values of $p_T$.

\section{Conclusions}

Studies of quarkonium production provide a unique laboratory in which to
understand the interplay between the perturbative and nonperturbative aspects
of QCD. They also provide an important testing ground for techniques for
perturbative calculations of high-energy production cross sections. 

At present, the primary method for theoretical calculations of
quarkonium production is the NRQCD factorization approach. NRQCD
factorization provides a systematic framework for computing quarkonium
production. However, there is no proof of NRQCD factorization for
quarkonium production beyond two loops. Furthermore, in spite of
considerable theoretical and experimental progress over the last decade,
a definitive test of NRQCD factorization has not yet been achieved.

Large corrections at NLO in $\alpha_s$ to inclusive quarkonium
production are now believed to be understood in terms of kinematic
enhancements. This understanding suggests that, for the color-octet
production channels, which are believed to be much more important than
the color-singlet production channels, there will be no large
kinematic enhancements beyond NLO. If that is the case, then the 
existing NLO 
calculations of quarkonium
production should at least be qualitatively correct. However, it seems
likely that resummation of large logarithms will be needed in order to
bring theory into good agreement with experiment at the large values of
quarkonium $p_T$ that will be probed at the LHC.

The predictions of the color-singlet model fail to describe the data. In
contrast, the predictions of NRQCD factorization are in agreement with
most of the inclusive production data. There are several notable
exceptions: (1) $J/\psi$ and $\Upsilon(2S)$ polarization at the
Tevatron, (2) the $e^+e^-\to J/\psi +X(\hbox{non}-c\bar c)$ total cross
section, and (3) $J/\psi$ production in $\gamma\gamma$ scattering at
LEP~II.  There are experimental discrepancies that cast some doubt on
the polarization measurements and on the measurement of the $e^+e^-\to
J/\psi +X(\hbox{non}-c\bar c)$ total cross section. The measurement of
$J/\psi$ production in $\gamma\gamma$ collisions has large experimental
uncertainties. Furthermore, even if NRQCD factorization is established 
at high $p_T$, it might not hold at the low values of $p_T$ at which 
this measurement was made.

The $J/\psi$ cross sections measured at the Tevatron and HERA, the
$J/\psi$ polarization measured by the CDF Collaboration in Run~II, and
NRQCD factorization seem to be mutually incompatible. Furthermore, the
new polarization prediction of Gong {\it et al.}\ \cite{Gong:2012ug},
which appeared after this talk was presented, suggests that the $J/\psi$
cross sections measured by the CDF and LHCb collaborations, the
$J/\psi$ polarization measured by the CDF Collaboration in               
Run~II, and NRQCD factorization are mutually incompatible. Some possible
explanations for these incompatibilities are: (1) the CDF Run~II
$J/\psi$ polarization measurement is wrong; (2) NRQCD factorization
fails at the low values of $p_T$ at which the HERA measurements were
made; (3) additional corrections to the theoretical predictions are
needed, such as those from higher orders in $v$; (4) NRQCD factorization
is incorrect.

Further advances in the theory of quarkonium production would help to
sort out the various possibilities. These include a proof or a
disproof of NRQCD factorization, calculations of corrections of higher
order in $\alpha_s$ and/or resummations of large logarithms of
$p_T^2/m_c^2$ for the fragmentation processes that are dominant at large
$p_T$, and calculations of the rates of additional production processes
that could help to pin down the NRQCD LDMEs and to provide further tests
of the production mechanisms.

Additional experimental measurements are essential to the process
of gaining an understanding of the quarkonium production mechanisms.
Experimental efforts that would be of immediate value are (1)
measurements $J/\psi$ {\it direct} production cross sections at high
$p_T$; (2) measurements of all three $J/\psi$ polarization parameters in
different frames for {\it direct} production at high $p_T$; (3)
measurements of $\chi_{cJ}$ cross sections and polarizations at high
$p_T$; (4) measurements of $\Upsilon$ {\it direct} production cross
sections at high $p_T$; (5), measurements of additional high-$p_T$
quarkonium production cross sections, such as the $J/\psi +\hbox{jet}$
cross section at high $p_T$. 

Since NRQCD factorization, if it is correct, likely holds only for $p_T$
substantially greater than $m_Q$, it is essential that all measurements
be made at the highest possible values of $p_T$. It is also very
important to make measurements of direct, rather than prompt, cross
sections. The feeddown contributions in prompt cross sections greatly
complicate the theoretical analyses, which may not be under good control
even in the simpler case of direct production. The measurements of
$\chi_{cJ}$ cross sections and polarizations might be particularly
interesting because the theory is more constrained: Only two NRQCD LDMEs
enter at the leading non-trivial order in $v$. $\Upsilon$ cross sections
and polarizations provide another important test of NRQCD  factorization
because the lower value of $v^2$ in the $\Upsilon(1S)$, in comparison
with the $J/\psi$, means that the $\Upsilon(1S)$ and $J/\psi$ LDMEs have
different relative sizes. It is clear that measurements of quarkonium
production processes and polarizations that take advantage of the high
$p_T$ reach and high luminosity of the LHC will play a large role in
advances in our understanding of quarkonium production mechanisms in the
next few years.
\vfill\eject

\Acknowledgements

I thank Mathias Butensch\"on, Kuang-Ta Chao, Bernd Kniehl, and Jian-Xiong 
Wang for many useful discussions. 
I also thank Jungil Lee, Hee Sok Chung, and Mathias Butensch\"on for
providing some of the figures that were used in this talk.
This work was supported by the U.~S.~Department of Energy, Division of
High Energy Physics, under Contract No.\ DE-AC02-06CH11357.


\begin{thebibliography}{99}


\bibitem{Bodwin:1994jh} 
  G.~T.~Bodwin, E.~Braaten and G.~P.~Lepage,
  Phys.\ Rev.\ D {\bf 51}, 1125 (1995)
  [Erratum-ibid.\ D {\bf 55}, 5853 (1997)].

\bibitem{Brambilla:2004wf} 
  N.~Brambilla {\it et al.}  [Quarkonium Working Group Collaboration],
  hep-ph/0412158.

\bibitem{Brambilla:2010cs} 
  N.~Brambilla, S.~Eidelman, B.~K.~Heltsley, R.~Vogt, G.~T.~Bodwin, 
  E.~Eichten, A.~D.~Frawley and A.~B.~Meyer {\it et al.},
  Eur.\ Phys.\ J.\ C {\bf 71}, 1534 (2011).

\bibitem{Kang:2011zza} 
  Z.~-B.~Kang, J.~-W.~Qiu and G.~Sterman,
  Nucl.\ Phys.\ Proc.\ Suppl.\  {\bf 214}, 39 (2011).
                 
\bibitem{Kang:2011mg} 
  Z.~-B.~Kang, J.~-W.~Qiu and G.~Sterman,
  Phys.\ Rev.\ Lett.\  {\bf 108}, 102002 (2012).

\bibitem{Kartvelishvili:1978id} 
  V.~G.~Kartvelishvili, A.~K.~Likhoded and S.~R.~Slabospitsky,
  Sov.\ J.\ Nucl.\ Phys.\  {\bf 28}, 678 (1978)
  [Yad.\ Fiz.\  {\bf 28}, 1315 (1978)].

\bibitem{Chang:1979nn} 
  C.~-H.~Chang,
  Nucl.\ Phys.\ B {\bf 172}, 425 (1980).

\bibitem{Berger:1980ni}
  E.~L.~Berger and D.~L.~Jones,
  Phys.\ Rev.\  D {\bf 23}, 1521 (1981).

\bibitem{Baier:1981uk} 
  R.~Baier and R.~Ruckl,
  Phys.\ Lett.\ B {\bf 102}, 364 (1981).

\bibitem{Baier:1983va} 
  R.~Baier and R.~Ruckl,
  Z.\ Phys.\ C {\bf 19}, 251 (1983).

\bibitem{Fritzsch:1977ay}                
  H.~Fritzsch,                                                
  Phys.\ Lett.\  B {\bf 67}, 217 (1977).                

\bibitem{Halzen:1977rs}
  F.~Halzen,
  Phys.\ Lett.\  B {\bf 69} 105, (1977).


\bibitem{Amundson:1995em}
  J.~F.~Amundson, O.~J.~P.~Eboli, E.~M.~Gregores and F.~Halzen,
  Phys.\ Lett.\  B {\bf 372}, 127 (1996).

\bibitem{Amundson:1996qr}
  J.~F.~Amundson, O.~J.~P.~Eboli, E.~M.~Gregores and F.~Halzen,
  Phys.\ Lett.\  B {\bf 390}, 323 (1997).

\bibitem{Yuan:2000qe}                                                    
  F.~Yuan and K.~T.~Chao,                                                
  Phys.\ Rev.\ Lett.\  {\bf 87}, 022002 (2001).

\bibitem{Baranov:2002cf}
  S.~P.~Baranov,
  Phys.\ Rev.\  D {\bf 66}, 114003 (2002).

\bibitem{Baranov:2007ay}
  S.~P.~Baranov and N.~P.~Zotov,
  JETP Lett.\  {\bf 86}, 435 (2007).

\bibitem{Baranov:2007dw}
  S.~P.~Baranov and A.~Szczurek,
  Phys.\ Rev.\  D {\bf 77}, 054016 (2008).

\bibitem{Nayak:2005rw} 
  G.~C.~Nayak, J.~-W.~Qiu and G.~F.~Sterman,
  Phys.\ Lett.\ B {\bf 613}, 45 (2005).

\bibitem{Nayak:2005rt} 
  G.~C.~Nayak, J.~-W.~Qiu and G.~F.~Sterman,
  Phys.\ Rev.\ D {\bf 72}, 114012 (2005).

\bibitem{Nayak:2006fm} 
  G.~C.~Nayak, J.~-W.~Qiu and G.~F.~Sterman,
  Phys.\ Rev.\ D {\bf 74}, 074007 (2006).

\bibitem{Artoisenet:2008fc}
  P.~Artoisenet, J.~M.~Campbell, J.~P.~Lansberg, F.~Maltoni and F.~Tramontano,
  Phys.\ Rev.\ Lett.\  {\bf 101}, 152001 (2008).

\bibitem{Abulencia:2007us} 
  A.~Abulencia {\it et al.}  [CDF Collaboration],
  Phys.\ Rev.\ Lett.\  {\bf 99}, 132001 (2007).

\bibitem{Acosta:2001gv} 
  D.~Acosta {\it et al.}  [CDF Collaboration],
  Phys.\ Rev.\ Lett.\  {\bf 88}, 161802 (2002).

\bibitem{Gong:2008sn} 
  B.~Gong and J.~-X.~Wang,
  Phys.\ Rev.\ Lett.\  {\bf 100}, 232001 (2008).

\bibitem{Ma:2010yw}                                                 
  Y.~-Q.~Ma, K.~Wang and K.~-T.~Chao,
  Phys.\ Rev.\ Lett.\  {\bf 106}, 042002 (2011).
                 
\bibitem{Butenschoen:2010rq}
  M.~Butenschoen and B.~A.~Kniehl,
  Phys.\ Rev.\ Lett.\  {\bf 106}, 022003 (2011).

\bibitem{Gong:2008ft} 
  B.~Gong, X.~Q.~Li and J.~-X.~Wang,
  Phys.\ Lett.\ B {\bf 673}, 197 (2009)
  [Erratum-ibid.\  {\bf 693}, 612 (2010)].

\bibitem{Campbell:2007ws}
  J.~M.~Campbell, F.~Maltoni and F.~Tramontano,
  Phys.\ Rev.\ Lett.\  {\bf 98}, 252002 (2007).

\bibitem{Ma:2010jj} 
  Y.~-Q.~Ma, K.~Wang and K.~-T.~Chao,
  Phys.\ Rev.\ D {\bf 84}, 114001 (2011).

\bibitem{bodwin-lee-frag}
  G.T.~Bodwin and J.~Lee, in preparation.

\bibitem{Ma:2010vd} 
  Y.~-Q.~Ma, K.~Wang and K.~-T.~Chao,
  Phys.\ Rev.\ D {\bf 83}, 111503 (2011).

\bibitem{Butenschoen:2011yh} 
  M.~Butenschoen and B.~A.~Kniehl,
  Phys.\ Rev.\ D {\bf 84}, 051501 (2011).

\bibitem{Butenschoen:2012qh} 
  M.~Butenschoen and B.~A.~Kniehl,
  Nucl.\ Phys.\ Proc.\ Suppl.\  {\bf 222-224}, 151 (2012).

\bibitem{Acosta:2004yw} 
  D.~Acosta {\it et al.}  [CDF Collaboration],
  Phys.\ Rev.\ D {\bf 71}, 032001 (2005).

\bibitem{Adloff:2002ex} 
  C.~Adloff {\it et al.}  [H1 Collaboration],
  Eur.\ Phys.\ J.\ C {\bf 25}, 25 (2002).

\bibitem{Aaron:2010gz} 
  F.~D.~Aaron {\it et al.}  [H1 Collaboration],
  Eur.\ Phys.\ J.\ C {\bf 68}, 401 (2010).
                        
\bibitem{Aaij:2011jh} 
  R.~Aaij {\it et al.}  [LHCb Collaboration],
  Eur.\ Phys.\ J.\ C {\bf 71}, 1645 (2011).

\bibitem{Aad:2011sp} 
  G.~Aad {\it et al.}  [ATLAS Collaboration],
  Nucl.\ Phys.\ B {\bf 850}, 387 (2011).

\bibitem{Adare:2009js} 
  A.~Adare {\it et al.}  [PHENIX Collaboration],
  Phys.\ Rev.\ D {\bf 82}, 012001 (2010).

\bibitem{Gong:2012ug} 
  B.~Gong, L.~-P.~Wan, J.~-X.~Wang and H.~-F.~Zhang,
  arXiv:1205.6682 [hep-ph].

\bibitem{Cho:1994ih} 
  P.~L.~Cho and M.~B.~Wise,
  Phys.\ Lett.\ B {\bf 346}, 129 (1995).

\bibitem{Bodwin:2005gg} 
  G.~T.~Bodwin, J.~Lee and D.~K.~Sinclair,
  Phys.\ Rev.\ D {\bf 72}, 014009 (2005).

\bibitem{Braaten:1999qk} 
  E.~Braaten, B.~A.~Kniehl and J.~Lee,
  Phys.\ Rev.\ D {\bf 62}, 094005 (2000).

\bibitem{Affolder:2000nn} 
  T.~Affolder {\it et al.}  [CDF Collaboration],
  Phys.\ Rev.\ Lett.\  {\bf 85}, 2886 (2000).

\bibitem{CDF-note-9966} CDF public note 9966.

\bibitem{Abazov:2008aa} 
  V.~M.~Abazov {\it et al.}  [D0 Collaboration],
  Phys.\ Rev.\ Lett.\  {\bf 101}, 182004 (2008).

\bibitem{Butenschoen:2011ks} 
  M.~Butenschoen and B.~A.~Kniehl,
  Phys.\ Rev.\ Lett.\  {\bf 107}, 232001 (2011).

\bibitem{Artoisenet:2009xh} 
  P.~Artoisenet, J.~M.~Campbell, F.~Maltoni and F.~Tramontano,
  Phys.\ Rev.\ Lett.\  {\bf 102}, 142001 (2009).

\bibitem{Butenschoen:2012px} 
  M.~Butenschoen and B.~A.~Kniehl,
  Phys.\ Rev.\ Lett.\  {\bf 108}, 172002 (2012).

\bibitem{Chao:2012iv} 
  K.~-T.~Chao, Y.~-Q.~Ma, H.~-S.~Shao, K.~Wang and Y.~-J.~Zhang,
  Phys.\ Rev.\ Lett.\  {\bf 108}, 242004 (2012).

\bibitem{Gong:2010bk} 
  B.~Gong, J.~-X.~Wang and H.~-F.~Zhang,
  Phys.\ Rev.\ D {\bf 83}, 114021 (2011).

\bibitem{Abelev:2011md} 
  B.~Abelev {\it et al.}  [ALICE Collaboration],
  Phys.\ Rev.\ Lett.\  {\bf 108}, 082001 (2012).

\bibitem{Chekanov:2009ad} 
  S.~Chekanov {\it et al.}  [ZEUS Collaboration],
  JHEP {\bf 0912}, 007 (2009).

\bibitem{Chatrchyan:2011kc} 
  S.~Chatrchyan {\it et al.}  [CMS Collaboration],
  JHEP {\bf 1202}, 011 (2012).

\bibitem{Pakhlov:2009nj} 
  P.~Pakhlov {\it et al.}  [Belle Collaboration],
  Phys.\ Rev.\ D {\bf 79}, 071101 (2009).

\bibitem{Zhang:2009ym} 
  Y.~-J.~Zhang, Y.~-Q.~Ma, K.~Wang and K.~-T.~Chao,
  Phys.\ Rev.\ D {\bf 81}, 034015 (2010).

\bibitem{Aubert:2001pd} 
  B.~Aubert {\it et al.}  [BABAR Collaboration],
  Phys.\ Rev.\ Lett.\  {\bf 87}, 162002 (2001).

\bibitem{Abdallah:2003du} 
  J.~Abdallah {\it et al.}  [DELPHI Collaboration],
  Phys.\ Lett.\ B {\bf 565}, 76 (2003).

\end{thebibliography}
\end{document}